\documentclass[a4paper,fleqn]{cas-sc}
\usepackage[authoryear,longnamesfirst]{natbib}

\usepackage{cas-common}
\usepackage{amsmath}
\usepackage{soul}
\usepackage{xcolor}
\usepackage[T1]{fontenc}
\usepackage{url}
\usepackage{amssymb}
\usepackage{multirow}
\usepackage{booktabs}
\usepackage{pifont}
\usepackage{graphicx}
\usepackage{longtable}
\usepackage{diagbox}
\usepackage{placeins}
\usepackage{amsthm}  % for theorem environments

\newtheoremstyle{mystyle} % name of the style
  {10pt} % Space above
  {10pt} % Space below
  {\itshape} % Body font
  {} % Indent amount
  {\bfseries} % Theorem head font
  {.} % Punctuation after theorem head
  {.5em} % Space after theorem head
  {} % Theorem head spec

\theoremstyle{mystyle}

\newtheorem{theorem}{Theorem}[section]

\newtheorem{mf}[theorem]{Main finding}

\def\tsc#1{\csdef{#1}{\textsc{\lowercase{#1}}\xspace}}
\tsc{WGM}
\tsc{QE}
\tsc{EP}
\tsc{PMS}
\tsc{BEC}
\tsc{DE}

\begin{document}
\let\WriteBookmarks\relax
\def\floatpagepagefraction{1}
\def\textpagefraction{.001}

\shorttitle{}
\shortauthors{Y. Huang et~al.}
\title [mode = title]{Uncovering key predictors of high-growth firms via explainable machine learning}

\author[1]{Yiwei Huang}
\credit{Conceptualization, Data curation, Formal analysis, Investigation, Methodology, Resources, Software, Validation, Visualization, Writing -- original draft, Writing -- review \& editing}

\author[2]{Shuqi Xu}
\credit{Conceptualization, Methodology, Funding acquisition, Supervision, Writing -- original draft, Writing -- review \& editing}

\author[3]{Linyuan L\"u} \cormark[1] \ead{linyuan.lv@ustc.edu.cn}
\credit{Conceptualization, Funding acquisition, Supervision, Project administration}

\author[4]{Andrea Zaccaria} \cormark[1] \ead{andrea.zaccaria@cnr.it}
\credit{Conceptualization, Methodology, Funding acquisition, Supervision, Project administration, Writing -- original draft, Writing -- review \& editing}

\author[5]{Manuel Sebastian Mariani}
\credit{Conceptualization, Methodology, Investigation, Funding acquisition, Supervision, Project administration, Writing -- original draft, Writing -- review \& editing}

\affiliation[1]{organization={Institute of Fundamental and Frontier Sciences}, addressline={University of Electronic Science and Technology of China}, city={Chengdu}, postcode={611731}, country={China}}

\affiliation[2]{organization={Institute of Dataspace}, addressline={Hefei Comprehensive National Science Center}, city={Hefei}, postcode={230088}, country={China}}

\affiliation[3]{organization={School of Cyber Science and Technology}, addressline={University of Science and Technology of China}, city={Hefei}, postcode={230026}, country={China}}

\affiliation[4]{organization={Istituto dei Sistemi Complessi (ISC-CNR)}, addressline={UoS Sapienza}, city={Rome}, postcode={I-00185}, country={Italy}}

\affiliation[5]{organization={URPP Social Networks}, city={Zurich}, postcode={CH-8050}, country={Switzerland}}

\cortext[1]{Corresponding authors.}

\begin{abstract}
Predicting high-growth firms has attracted increasing interest from the technological forecasting and machine learning communities.
Most existing studies primarily utilize financial data for these predictions.
However, research suggests that a firm's research and development activities and its network position within technological ecosystems may also serve as valuable predictors.
To unpack the relative importance of diverse features, this paper analyzes financial and patent data from $5,071$ firms, extracting three categories of features: financial features, technological features of granted patents, and network-based features derived from firms' connections to their primary technologies.
By utilizing ensemble learning algorithms, we demonstrate that incorporating financial features with either technological, network-based features, or both, leads to more accurate high-growth firm predictions compared to using financial features alone.
To delve deeper into the matter, we evaluate the predictive power of each individual feature within their respective categories using explainable artificial intelligence methods.
Among non-financial features, the maximum economic value of a firm's granted patents and the number of patents related to a firms' primary technologies stand out for their importance.
Furthermore, firm size is positively associated with high-growth probability up to a certain threshold size, after which the association plateaus.
Conversely, the maximum economic value of a firm's granted patents is positively linked to high-growth probability only after a threshold value is exceeded.
These findings elucidate the complex predictive role of various features in forecasting high-growth firms and could inform technological resource allocation as well as investment decisions.
\end{abstract}


\maketitle

\section{Introduction}

Research on high-growth firms (HGFs) has substantially grown over the past decade~\citep{jansen2023scaling}.
HGFs play a key role in job creation, attracting investments, and advancing technology~\citep{weinblat2018forecasting}, which makes the HGF prediction not only a theoretically important problem but also a key quantitative tool to inform economic strategies and policies.
To predict HGFs, most studies to date relied on features extracted from firms' financial information~\citep{coad2020catching,chae2024search}.
Recent studies have included aggregated features related to research and development (R\&D) activities, such as R\&D expenditure~\citep{segarra2014high} and the number of patents~\citep{virtanen2019predicting,guzman2020state,chae2024search}, yet most of them neglect the heterogeneous characteristics and impacts of the main outputs of R\&D, namely patents.
In parallel, fueled by recent advances in network science, the economic complexity field has recently developed and validated several network-based indicators to predict the future development of countries~\citep{tacchella2012new,tacchella2018dynamical,mariani2019nestedness,ye2022forecasting}.
Testing and adapting these methods from macro-prediction (at the scale of national economies) to the micro level (at the scale of individual firms) has remained elusive so far.

As a result, there lacks a systematic comparison regarding the importance of financial, technological, and network-based features in predicting HGFs.
In particular, several questions remain open: Do patent-based features improve the predictability of HGFs above and beyond past financial performance?
Among patent-based features, is the economic or technological value of a firm's granted patents a stronger predictor of high growth?
Can features built from the firms' technological networks improve the predictability of HGFs?
If yes, which features are the most important for accurate predictions?
To address these questions, we analyze the lifecycles of $5,071$ firms for which we could match their patenting activities (extracted from the USPTO dataset) with their financial information (extracted from the Orbis dataset).
We include three classes of features: financial features, technological features (i.e., features based on the characteristics of a firm's granted patents~\citep{wu2023identifying,xu2024citations}), and network-based features (i.e., features extracted from the bipartite networks that connect the firms with their primary technologies~\citep{pugliese2019coherent,straccamore2022will,kim2023technological}).
By employing ensemble learning algorithms, we leverage these features to predict the top-performing firms in the growth of three key financial variables: the number of employees, turnover, and net income.

We find that incorporating financial features with either technological, network-based features, or both, leads to more accurate HGF predictions compared to using financial features alone.
The best-performing model includes both technological and network-based features for almost all the target variables considered here, which indicates that both classes of features play a crucial role in HGF prediction.
We employ eXplainable Artificial Intelligence (XAI) methods, including Gini importance, SHapley Additive exPlanations (SHAP), and Partial Dependence Plots (PDP), to analyze the influence of individual features on the predictive performance.
Among the technological features, the maximum economic value of a firm's granted patents tends to be the most important predictor.
In particular, it is more important than citation-based indicators for all the analyzed target variables.
The probability of a firm achieving high growth is positively associated with the maximum economic value of its patents, but only after a threshold value is exceeded.
Among the network-based features, the number of patents related to a firm's primary technologies consistently holds the highest importance, whereas the Fitness-Complexity algorithm---originally developed to predict national development~\citep{tacchella2012new}---does not exhibit high importance scores.

To the best of our knowledge, this paper is the first work integrating patent attributes, network analysis methods, and explainable artificial intelligence methods into HGF prediction.
In summary, we
\begin{itemize}
    \item leverage the inherent attributes of patents to measure firms' technological capabilities
    \item consider the impact of the features of firms' patenting activities on high-growth predictions
    \item use network analysis methods to measure firms' roles and positions within broader technological ecosystems
    \item integrate XAI methods into predictive models, which uncover the contribution of firms' features to achieving high growth.
\end{itemize}

The explainability of our framework allows us to provide managerial implications for strategic decision-making on firm development.

\section{Related literature}

\subsection{R\&D activities and firm growth}

The nature of firm growth is heterogeneous, complex, and dynamic, influenced by both internal characteristics and the external environment~\citep{audretsch2014firm}.
Besides, the process of R\&D activities is full of risk and uncertainty~\citep{delmar2003arriving}.
Even successful innovations require considerable investments to translate into tangible economic performance~\citep{coad2008innovation}.
These complexities lead to varied impacts of R\&D activities on firm growth, with extensive studies exploring the relationship between innovation and firm growth~\citep{audretsch2014firm,segarra2014high}.
\cite{coad2016innovation} observed that the impact of R\&D on growth is more predictable for mature firms, whereas young firms exhibit a riskier return profile from R\&D investments, including the potential for both larger benefits and greater losses.
\cite{leyva2022inverted} revealed the negative moderating effect of firm age on the relationship between environmental innovation and firm performance, as it is easier for young firms to benefit from environmental innovations than for mature firms.
As for firm size, \cite{demirel2012innovation} found that large pharmaceutical companies may experience a negative effect on growth from R\&D activities, while small firms can see a positive growth impact only if they persistently pursue innovation activities.
The industry context is another key factor.
\cite{garcia2012research} observed that only high-technology industries benefit from their R\&D investment, which is consistent with the findings that R\&D shows a higher positive impact on growth for manufacturing industries than for service industries~\citep{segarra2014high}.
Overall, the impact of R\&D activities on firm growth is significant and complex, which has stimulated studies leveraging R\&D information for predicting firms' future performance.

In recent years, there has been a substantial increase in research on high-growth firms~\citep{jansen2023scaling}.
The increasing focus on HGFs stems from their significant role in job creation, investment attraction, and technological advancements~\citep{weinblat2018forecasting}.
This makes the prediction of HGFs not only a theoretical interest but also a practical necessity for economic strategies and policy development.
According to the comprehensive review articles by~\cite{coad2020catching} and~\cite{chae2024search}, previous studies on HGF prediction have evolved in both the considered features and predictive models.

Given that most studies define HGFs based on firms' financial metrics (see Section~\ref{sec:target} for details), it is common to extract features from firms' financial information.
Most financial features relate to firms' internal characteristics such as the number of employees, turnover, and profits~\citep{MIYAKAWA2017,virtanen2019predicting,coad2020catching,chae2024search}.
Beyond financial variables, recent studies increasingly include features based on firms' R\&D activities.
\cite{lee2014holds} and \cite{goedhuys2016high} introduced relevant features to indicate whether a firm is active in R\&D.
\cite{simbana2019key} measured firms' R\&D intensity as the ratio between intangible assets to total assets, while \cite{segarra2014high} considered the expenditure on internal and external R\&D activities.
In addition, the number of patents has also been utilized in recent studies~\citep{virtanen2019predicting,guzman2020state,chae2024search}.

However, both R\&D expenditure and patent counts reflect the scale of R\&D activities while ignoring their inherent attributes, which could be misleading to measure firms' technological capabilities~\citep{coombs2006measuring}.
Patents, as the main outputs of innovation activities, offer valuable insights into the assessment of firms' technological capabilities and their potential for future growth.
For instance, the number of citations received by a given patent (i.e., its citation count) reflects the patent's technological impact in the industry~\citep{wu2023identifying}.
The movement of stock prices following a patent's grant can reflect its economic value~\citep{kogan2017technological}.
These patent-based features help capture and anticipate different aspects of firms' technological performance~\citep{xu2024citations}.
However, most of them are not included in the existing HGF prediction studies.
Besides, current studies focus on firms' individual features and neglect the relative position of firms within the context of economic and technological ecosystems.
Network analysis methods can bridge this gap by building network models, focusing on how relationships and positions within these networks affect firms' innovation and growth~\citep{chuluun2017firm,jeude2019multilayer}.

Our work contributes to this growing literature by leveraging the attributes of a given firm's patents to predict high growth.
This approach reveals the most important features of firms' patenting activities that predict high growth.
\cite{xu2024citations} found that the economic value of patents is a strong predictor of a firm's ability to produce high-value patents in the future.
Here we extend this result by showing that the maximum economic value of a firm's granted patents is also a stronger predictor of future high growth.
Besides, we resort to network analysis methods to measure firms' roles and positions within broader technological ecosystems, and systematically compare the predictive power of diverse network-based features.

\subsection{Machine learning models for high-growth firm prediction}

As for the predictive models used in HGF prediction, previous studies commonly utilized regression models, such as Probit regression~\citep{lopez2012makes,arrighetti2013assessing,bjuggren2013high,daunfeldt2014economic,lee2014holds,goedhuys2016high,bianchini2017does,megaravalli2018predicting,pereira2018impact}, Quantile regression~\citep{sampagnaro2013identifying,segarra2014high,simbana2019key}, Linear regression~\citep{holzl2014persistence,moschella2019persistent}, Least Absolute Shrinkage and Selection Operator (LASSO)~\citep{mckenzie2019predicting,coad2020catching,chae2024search}, and Logistic regression~\citep{zekic2016predicting,guzman2020state}.
With the development of information technology, machine learning (ML) models have emerged as powerful tools for predicting HGFs, such as Support Vector Machines (SVM)~\citep{mckenzie2019predicting}, Artificial Neural Network (ANN)~\citep{zekic2016predicting}, and random forest~\citep{MIYAKAWA2017,weinblat2018forecasting,virtanen2019predicting}.
Compared to traditional regression models, ML models excel in processing complex and non-linear relationships among features.

However, while these ML models show superior predictive performance in predicting HGFs, the complexity of their internal mechanisms makes the predictions difficult to explain, leading to the ``black box'' problem~\citep{adadi2018peeking,minh2022explainable}.
This lack of transparency and interpretability becomes a critical issue for external users such as policymakers, managers, and investors, who focus not only on the prediction accuracy but also on the key factors driving a firm's potential for high growth~\citep{zhang2022explainable}.
Although random forest models offer some degree of interpretability through feature importance, this method doesn't explain how specific features influence the prediction in terms of direction (positive or negative) or magnitude at the individual prediction level.

To solve the black box problem, the field of eXplainable Artificial Intelligence (XAI) has developed many techniques to enhance model interpretability, among which model-agnostic approaches are the most popular ones due to their flexibility to be applied to any ML model~\citep{adadi2018peeking}.
According to the interpretability scope, model-agnostic approaches can be further classified into global and local model-agnostic approaches~\citep{molnar2022}.
The former class provides an overall understanding of a ML model's decision logic that applies across the entire dataset, while the latter class explains the model's decision on a single sample.
So far, XAI methods have been applied in a variety of firm studies, such as financial distress prediction~\citep{zhang2022explainable,jiang2023mining,che2024predicting,hajek2024corporate}, venture capital financing~\citep{zbikowski2021machine,niculaescu2023venture}, and credit risk management~\citep{bussmann2021explainable,wang2023qualitatively}, but the application in HGF prediction still remains unexplored.

To the best of our knowledge, this paper is the first work integrating patent attributes, network analysis methods, and XAI methods into HGF prediction.
We integrate XAI methods into predictive models, which interpret the contribution of firms' features on achieving high growth.
This interpretation also gives managerial implications for strategic decision-making on firm development.

\subsection{Economic complexity}

Economic complexity~\citep{hidalgo2021economic,balland2022new} is a framework that studies economic development by adopting methods from network science, statistical physics, and machine learning. Traditionally, economic complexity metrics, such as the Fitness-Complexity algorithm~\citep{tacchella2012new}, are computed from international trade data and show high predictive performance at \textit{country} level: for instance, \cite{tacchella2018dynamical} used the Fitness-Complexity algorithm to predict GDP growth better than state-of-the-art forecasts by the International Monetary Fund, while \cite{tacchella2023relatedness} and \cite{albora2023product} compared machine learning with network-based algorithms to forecast which country will start exporting a given product.
More recently, these approaches have also been applied to industrial and patenting activities at the \textit{firm} level.
In particular, by adopting machine learning techniques, \cite{albora2022machine} and \cite{straccamore2022will} forecast, respectively, the next product exported and the next technology patented by a given firm.
\cite{arsini2023prediction} investigated the patenting activities of firms to forecast Mergers and Acquisitions.
However, the identification of the determinants of firms' growth, and the identification of the best-performing firms have remained elusive.
In this work, we find that the Fitness-Complexity algorithm, previously developed for countries, is not a strong predictor of firms' growth.
This confirms the conjectures by~\cite{laudati2023different}, who found that the application of this algorithm at the firm level might be ineffective because of the modular, in-block nested structure of the firm-product bipartite network.
By contrast, here we show that compared to the Fitness-Complexity algorithm, a more important network-based predictor of growth is simply the number of patents related to a firm's primary technologies.

\section{Data}

\subsection{Data collection}\label{sec:collection}

The data collection process consists of two steps: collecting firms' patenting activities and collecting firms' financial performance.
To obtain a unified set of research objects (firms), we start from Michael Woeppel's patent database\footnote{\url{https://www.mikewoeppel.com/data}}~\citep{stoffman2022small}, in which each patent is a utility patent granted by the United States Patent and Trademark Office (USPTO), and the assignees of these patents are firms listed in the stock market.
Firms in the dataset are identified by a unique permanent identifier named ``permco'', which is assigned by the Center for Research in Security Prices (CRSP).
In total, we have collected a patent dataset composed of $8,673$ firms and their $2,757,366$ issued utility patents filed between $1975$ and $2020$.

\begin{figure*}[pos=hbt]
    \centering
    \caption{Schematic visualization of the process we adopt to collect and organize the patent and financial datasets.
    More details of the matching process are provided in Appendix~\ref{sec:appendix_matching}.}
    \includegraphics[width=12cm]{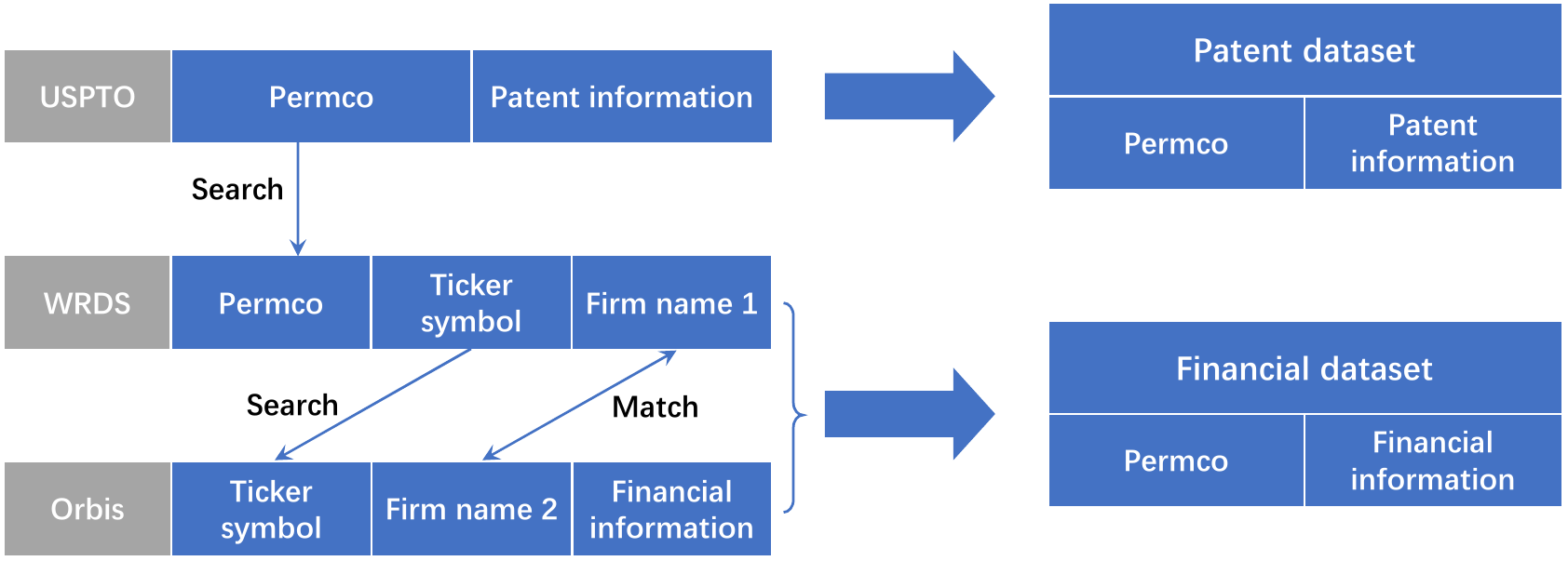}
    \label{fig:1}
\end{figure*}

To collect the financial performance of firms, we utilize Wharton Research Data Services\footnote{\url{https://wrds-www.wharton.upenn.edu}} (WRDS) to retrieve firms' names and ticker symbols (to be associated with the permco).
Then we conduct a search using the ticker symbols to obtain the financial performance of involved firms in the Orbis database\footnote{\url{https://www.bvdinfo.com}}.
Considering that a ticker symbol can be reassigned and thus reused by different firms, we employ the fuzzy matching method\footnote{\url{https://github.com/seatgeek/fuzzywuzzy}}~\citep{cohen2011fuzzywuzzy} to calculate the matching scores between the firm names in the Orbis and WRDS data.
This allows us to estimate if these data correspond to the same firm.
Further details of data matching are described in Appendix~\ref{sec:appendix_matching}.
After successfully matching $5,071$ firms, we have collected a financial dataset that includes firms' yearly financial statements between $1994$ and $2023$, in terms of the number of employees, turnover, and net income.
Figure~\ref{fig:1} shows the process of matching the patent dataset and financial dataset.

\subsection{Data processing} \label{sec:process}

\begin{figure*}[pos=hbt]
    \centering
    \caption{Overview of our 7-year time window organization of the data.}
    \includegraphics[width=11cm]{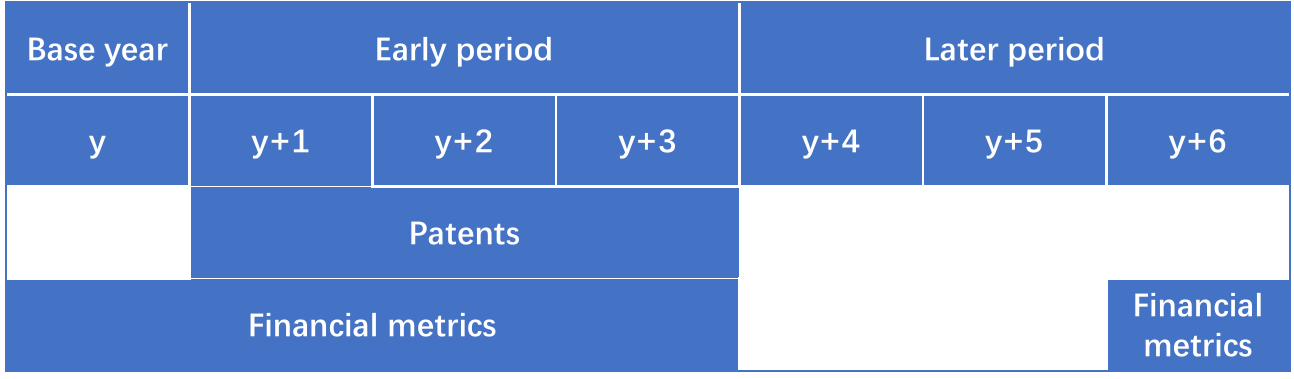}
    \label{fig:2}
\end{figure*}

Given the yearly fluctuation of firms' financial performance, we extract a 7-year time window from a given base year $y$ to the subsequent $6$ years ($y+1, y+2, \dots, y+6$).
In this time window, we classify the period from $y+1$ to $y+3$ as the \textit{early period} and the period from $y+4$ to $y+6$ as the \textit{later period}.
First, we extract patents filed from year $y+1$ to $y+3$, which correspond to firms' patenting activities within the early period.
We remove firms that have not filed patents during this period.
The remaining firms and their patents will constitute our patent dataset.
Then, we extract firms' financial metrics in five different years: $y, y+1, y+2, y+3$ and $y+6$.
In the analysis presented in the main text, we preserve only those firms that have available financial metrics for all of these five years. This will constitute our financial dataset.
To exclude a potential survivor bias introduced by this data processing, we replicate our experiments using less conservative data filtering criteria.
The results, presented in Section~S3 of the supplementary material (see Appendix~\ref{appendix:sm}), confirm the robustness of our findings.
Figure~\ref{fig:2} provides an overview of the organization of our data in 7-year time windows.

After extracting data for a specific time window, the next step is to merge the two extracted datasets into the final dataset.
We utilize permco as the identifier to incorporate the patent information for the corresponding firms in the extracted financial dataset.
For firms that have no corresponding patent information, we make the assumption that these firms have zero patenting activity within the given time window.
In the main text, our focus is the time window from $2000$ to $2006$, i.e. base year $y$ is $2000$; in this case, the final dataset consists of $1,929$ firms.
To confirm the robustness of our results, we also consider different time windows; these analyses can be found in Section~S1 of the supplementary material (see Appendix~\ref{appendix:sm}).

\section{Methods} \label{sec:method}

In this section, we describe the variables and the methodologies involved in our investigation. In particular, the first subsection illustrates our target variables: namely, how we identify HGFs in the later period. Then we describe the feature variables: the financial, technological, and network-based characteristics of the firms.
These feature variables are computed in the early period and are used to predict whether a firm will achieve high growth or not. The last two subsections are devoted, respectively, to the description of the ML classifiers and the evaluation metrics we use to quantify the predictive performance.

\subsection{Target variables} \label{sec:target}

HGF prediction has been a focus of scholarly interest, yet there exists no shared definition of what an HGF is.
This lack of consensus primarily arises from the direct influence of chosen growth metrics, measurement methods, and classification criteria on the identification of HGFs~\citep{delmar2003arriving,coad2014high,chae2024search}.
Indeed, different combinations of these elements capture various aspects of firms' performance improvement.

The definition proposed by Eurostat and OECD~\citep{eurostat2007eurostat} has been widely adopted in previous studies~\citep{pereira2018impact,coad2020catching,chae2024search}, which identifies HGFs as those starting with at least $10$ employees and achieving an average annualized growth rate exceeding $20\%$ in the number of employees or turnover over a three-year period.
The number of employees and turnover are the most commonly used metrics~\citep{daunfeldt2014economic,coad2020catching}, since they act as complementary metrics reflecting firm's labor market dynamics and industrial dynamics, respectively~\citep{moschella2019persistent}.
However, the calculation of the relative growth rate introduces a bias towards smaller firms due to their potential for rapid percentage increases from a small base~\citep{coad2014high}.
In contrast, the calculation of an absolute growth quantity, which measures the direct change in the considered metrics, tends to favor larger firms, as their larger base allows for substantial numerical changes~\citep{coad2014high}.
To reduce the impact of firm size, we will adopt a popular growth measurement method, the Birch-Schreyer index~\citep{schreyer2000high}, which combines the relative and absolute change~\citep{lopez2012makes,holzl2014persistence,weinblat2018forecasting,long2019becoming,eklund2020some,de2021spillovers}.
Another limitation of the Eurostat-OECD definition is its reliance on an absolute growth rate threshold (i.e. $20\%$) as the classification criterion, which overlooks the variations in growth rates across different industries~\citep{sampagnaro2013identifying}.

Given the above insights from the review of the available data and HGF definitions, in this work we perform a comprehensive analysis that considers five growth indicators.
Our analysis combines three key financial metrics, i.e. the number of employees, turnover, and net income, with two growth measurement methods, i.e. the absolute growth quantity (AG) and the Birch-Schreyer index (BSI), as shown in Table~\ref{tab:1}.
Note that we exclude the Birch-Schreyer index for net income, due to the presence of negative net income values which undermines the interpretation of a relative growth.
As for the classification criteria, we define HGFs as those firms ranking in the top $10\%$ within their respective industries for each growth indicator.
Such a criterion avoids the subjectivity inherent in using an absolute threshold and addresses the variations in growth indicators across different industries.
The industries are identified based on the first $2$ digits of firms' NAICS (North American Industry Classification System) codes\footnote{\url{https://www.naics.com/six-digit-naics}}.
In summary, the target variables in our analysis are binary values indicating whether a firm is considered an HGF or not, in terms of each growth indicator.

\begin{table}[pos=hbt]
\centering
\caption{List of growth indicators. $X_{y}$ refers to the financial metric value in year $y$.}
\resizebox{\linewidth}{!}{
\begin{tabular}{p{4cm}p{4cm}p{12cm}}
\hline
Notation & Formula  & Description \\ \hline
AG (Employee)           & $X_{y+6} - X_{y+3}$ & The absolute growth quantity of the number of employees in the later period.\\
AG (Turnover)           & $X_{y+6} - X_{y+3}$ & The absolute growth quantity of turnover in the later period.\\
AG (Net\_income)        & $X_{y+6} - X_{y+3}$ & The absolute growth quantity of net income in the later period.\\
BSI (Employee)         & $(X_{y+6} - X_{y+3})(X_{y+6}/X_{y+3})$ & The Birch-Schreyer index of the number of employees in the later period.\\
BSI (Turnover)         & $(X_{y+6} - X_{y+3})(X_{y+6}/X_{y+3})$ & The Birch-Schreyer index of turnover in the later period.\\
\hline
\end{tabular}}
\label{tab:1}
\end{table}

\subsection{Financial features} \label{sec:financial}

Financial features capture firms' financial performance within the early period, including the quantity recorded in the last year, the average values over the early period, and the growth indicators of the financial metrics.
The explicit formulation of the financial features is listed in Table~\ref{tab:2}.

\begin{table}[pos=hbt]
\centering
\caption{List of the financial features we use as predictors to identify high-growth firms.}
\resizebox{\linewidth}{!}{
\begin{tabular}{p{4cm}p{4cm}p{12cm}}
\hline
Notation & Formula  & Description \\ \hline
Employee\_last      & $X_{y+3}$ & The number of employees in the last year of the early period.\\
Turnover\_last      & $X_{y+3}$ & Turnover in the last year of the early period.\\
Net\_income\_last   & $X_{y+3}$ & Net income in the last year of the early period.\\
Employee\_average     & $(X_{y+1}+X_{y+2}+X_{y+3})/3$ & Average number of employees in the early period.\\
Turnover\_average     & $(X_{y+1}+X_{y+2}+X_{y+3})/3$ & Average turnover in the early period.\\
Net\_income\_average  & $(X_{y+1}+X_{y+2}+X_{y+3})/3$ & Average net income in the early period.\\
AG (Employee)           & $X_{y+3} - X_{y}$ & The absolute growth quantity of the number of employees in the early period.\\
AG (Turnover)           & $X_{y+3} - X_{y}$ & The absolute growth quantity of turnover in the early period.\\
AG (Net\_income)        & $X_{y+3} - X_{y}$ & The absolute growth quantity of net income in the early period.\\
BSI (Employee)         & $(X_{y+3} - X_{y})(X_{y+3}/X_{y})$ & The Birch-Schreyer index of the number of employees in the early period.\\
BSI (Turnover)         & $(X_{y+3} - X_{y})(X_{y+3}/X_{y})$ & The Birch-Schreyer index of turnover in the early period.\\
\hline
\end{tabular}}
\label{tab:2}
\end{table}

\subsection{Technological features}\label{sec:tech_feature}

The technological features capture the information related to firms' patenting activities, including the number of patents, the number of patents' citations, the number of related technologies, and the value of patents.
Table~\ref{tab:3} lists all technological features.

Each patent contains one or more Cooperative Patent Classification (CPC) codes that identify the technology sectors to which it belongs.
Specifically, we identify these sectors using the first $4$ characters of the CPC codes present in the patent, which represent the \textit{related technologies} of the patent.
The number of related technologies of a firm is then defined as the total number of unique related technologies among its patents.
The value of patents can be quantified from two dimensions: the technological value and the economic value~\citep{xu2024citations}.
The technological value of patents is quantified by the number of citations they receive.
The economic value of patents is measured based on the movement of firms' stock price over a narrow time period following their issuance~\citep{kogan2017technological,stoffman2022small}.
However, both measurements can be influenced by the bias introduced by patent age, thus we employ age-normalized measures for technological and economic value~\citep{xu2024citations}.
These age-normalized measures take into account the ranking positions of patents based on their citation counts and stock returns, respectively, among the patents issued in the same year.
We utilize the minimum, maximum, and mean technological and economic values of patents to evaluate a firm's overall innovative capabilities.
The normalized technological value (NTV) of patent $i$ reads~\citep{xu2024citations}
$NTV_i = 1-r_i/N(t_i)$
where $N(t_i)$ is the number of patents issued in the same year $t_i$ as patent $i$, and $r_i$ denotes the ranking of $i$ by citation count among the $N(t_i)$ patents.
The patent with the highest citation count has $r_i=1$, while $r_i=N(t_i)$ corresponds to the one with the lowest citation count.
We assign the average rankings to the patents with the same citation count.
Thus, $0 \leq NTV_i < 1$.
Similarly, the normalized economic value (NEV) of patent $i$ reads~\citep{xu2024citations}
$NEV_i = 1-r_i/N(t_i)$,
where $r_i$ denotes the ranking of $i$ by economic value among the $N(t_i)$ patents.
Also, $0 \leq NEV_i < 1$.

\begin{table}[pos=hbt]
\centering
\caption{List of technological features.}
\resizebox{\linewidth}{!}{
\begin{tabular}{p{4cm}p{16cm}}
\hline
Notation & Description \\ \hline
No. patent     & The number of granted patents filed within the early period.\\
No. Technology       & The number of related technologies.\\
Cited\_num\_min & The minimum citation count of granted patents filed within the early period.\\
Cited\_num\_mean & The mean citation count of granted patents filed within the early period.\\
Cited\_num\_max & The maximum citation count of granted patents filed within the early period.\\
Tech\_value\_min & The minimum technological value of granted patents filed within the early period.\\
Tech\_value\_mean & The mean technological value of granted patents filed within the early period.\\
Tech\_value\_max & The maximum technological value of granted patents filed within the early period.\\
Eco\_value\_min & The minimum economic value of granted patents filed within the early period.\\
Eco\_value\_mean & The mean economic value of granted patents filed within the early period.\\
Eco\_value\_max & The maximum economic value of granted patents filed within the early period.\\
\hline
\end{tabular}}
\label{tab:3}
\end{table}

\subsection{Network-based features}

To make a more comprehensive evaluation of firms' patenting activities, we construct two networks: the firm-technology bipartite network and the industry-technology bipartite network.
The firm-technology network is an unweighted network where firms are linked with their \textit{primary technologies}.
We first calculate the relative number of filed patents for each firm-technology pair; after a suitable normalization, if this normalized value is higher than a threshold, the technology is defined as a primary technology for that firm.
In practice, this is done by computing the Revealed Comparative Advantage (RCA) index~\citep{balassa1965trade}.
Based on the firm-technology network, we build the industry-technology network by aggregating firms into industries according to their NAICS codes.
In this network, the weight of each industry-technology link represents the number of firms within that industry for which the technology is their primary technology.
However, spurious co-occurrences may arise, simply deriving from the diversification of firms' patenting activities and the ubiquity of very common technologies.
In order to take this into account and filter out these noisy links, we build randomized firm-technology networks using the the Bipartite Configuration Model (BiCM)\footnote{\url{https://github.com/mat701/BiCM}}~\citep{saracco2015randomizing,vallarano2021fast}, and for each of these random networks we build the corresponding industry-technology network.
By comparing the link weights of the empirical network with the distribution of the link weights from the randomized counterparts, we obtain the statistically significant industry-technology links~\citep{cimini2022meta}.
The construction of the two networks is detailed in Appendix~\ref{sec:appendix_network}.
Table~\ref{tab:4} lists all network-based features, which can be classified into three categories.

\subsubsection{Features based on firms' primary technologies}

When building the firm-technology network, we compute the RCA index for each firm-technology pair based on the number of patents.
We then connect firms to their primary technologies, which are identified as those with an RCA higher or equal to $1$.
Thus we can quantify the frequency and ratio of firms' patenting activities regarding their primary technologies, which reflect their technological leadership and strategic focus in primary areas.

\subsubsection{Features based on economic complexity methods}

The building of a firm-technology network allows the incorporation of economic complexity methods for analyzing and quantifying various aspects of firms' technological innovation.

The Fitness-Complexity algorithm~\citep{tacchella2012new} has been used to accurately forecast the future economic growth of countries~\citep{tacchella2018dynamical}, which suggests that it might accurately gauge an economic actor's capabilities.
In this algorithm, the fitness scores of countries and the complexity scores of products are assessed by the stationary state of a nonlinear iterative process~\citep{tacchella2012new}.
The core idea of this algorithm is that competitive countries export as many products as possible, compatibly with their constraints; at the same time, sophisticated products are only exported by highly competitive countries.
While the Fitness-Complexity equations have been originally introduced through heuristic considerations, it can be shown that they solve a long-known optimization problem where a suitable cost function penalizes exports of sophisticated products from low-competitiveness countries~\citep{mazzilli2024equivalence}.
Given the Fitness-Complexity method's proven ability to precisely quantify countries' competitiveness, we extend its application to the evaluation of firms' competitiveness within the firm-technology network, resulting in the calculation of firm-level technological fitness scores.
The mathematical formulation of the Fitness-Complexity method is detailed in Appendix~\ref{sec:fc}.

Coherent technological diversification (CTD)~\citep{pugliese2019coherent} is a measurement of the coherence of firms' technological capabilities, which exhibits a significant correlation with productive efficiency.
The core idea of CTD is quantifying the coherence between each of a firm's primary technology and its technological portfolio.
This involves the definitions of the proximity between technologies and, as a second step, the computation of the global coherence of firms' technological portfolios.
In this paper, we focus on three distinct definitions of technological proximity.
The first definition leverages information from patent applicants, defining proximity between two technologies from suitably normalized firm-level co-occurrences.
Various proximity measures exist; one of these is equivalent to computing the minimum of the pairwise conditional probabilities that a firm has one technology as the primary technology, given that it has another technology as the primary technology~\citep{hidalgo2007product,zaccaria2014taxonomy,pugliese2019coherent,hidalgo2021economic}.
The second definition relies on patent references~\citep{yan2017measuring}, where proximity is measured by the Jaccard similarity between the citation sets of patents in two technologies.
The third definition utilizes patent classifications~\citep{yan2017measuring}, defining proximity as the Jaccard similarity between the sets of patents in two technologies.
As for the coherence computation, we consider the sum and mean proximity between a technology and a firm's technological portfolio~\citep{pugliese2019coherent}.
The different combinations of technological proximity definitions and coherence computation methods result in $8$ different CTD metrics.
The details of all CTD metrics are discussed in Appendix~\ref{sec:ctd}.

\subsubsection{Features based on industries' related technologies}

In the industry-technology network, the link weights exhibit statistical significance compared to those in the randomized networks.
As a result, industries are connected to their significantly related technologies.
These connections enable us to quantify the frequency and proportion of firms' patenting activities within the related technologies of their respective industries, which reflects the degree of ``agreement'' between firms' patenting activities and their industrial classification.

\begin{table}[pos=hbt]
\centering
\caption{List of network-based features.}
\resizebox{\linewidth}{!}{
\begin{tabular}{p{3cm}p{5cm}p{12cm}}
\hline
Bipartite network & Notation & Description \\
\hline
\multirow{4}{*}{Firm-technology} & No. patent (Primary)          & The number of granted patents related to a firm's primary technologies.\\
& No. Technology (Primary)            & The number of a firm's primary technologies.\\
& Patent ratio (Primary)        & The ratio of granted patents related to a firm's primary technologies.\\
& Technology ratio (Primary)          & The ratio of a firm's primary technologies.\\
\cline{1-3}
\multirow{9}{*}{Firm-technology} & Fitness                  & A firm's fitness score measured by the Fitness-Complexity method. \newline See Appendix~\ref{sec:fc} for details.\\
& CTD (Applicant-TS-Sum)         &\multirow{8}{*}{\begin{tabular}[c]{@{}l@{}}Various measurements of a firm's coherent technological diversification (CTD), which\\ differ in their definitions on technological proximity and the computation of coherence\\ between a technology and a firm's technological portfolio.\\
See Appendix~\ref{sec:ctd} for details.\end{tabular}}\\
& CTD (Applicant-TS-Mean)         &\\
& CTD (Applicant-TN-Sum)         &\\
& CTD (Applicant-TN-Mean)         &\\
& CTD (Reference-Sum)        &\\
& CTD (Reference-Mean)        &\\
& CTD (Classification-Sum)  &\\
& CTD (Classification-Mean)  &\\
\hline
\multirow{4}{*}{Industry-technology}
& No. patent (Industry-related)          & The number of granted patents that belong to a firm's industry-related technologies.\\
& No. Technology (Industry-related)            & The number of a firm's industry-related technologies.\\
& Patent ratio (Industry-related)        & The ratio of granted patents that belong to a firm's industry-related technologies.\\
& Technology ratio (Industry-related)          & The ratio of a firm's industry-related technologies.\\
\hline
\end{tabular}}
\label{tab:4}
\end{table}

\subsection{Classifiers}

\subsubsection{Naïve classifier}

We refer to naïve classifiers as those utilizing only one feature to make predictions.
We introduced the financial, technological, and network-based features in the previous subsections.
For each of these features, we construct an individual naïve classifier, which predicts HGFs as those whose feature value ranks in the top $10\%$ within their respective industries.
By testing the performance of these classifiers, we can analyze independently the influence of each feature on identifying HGFs, as shown in Section~\ref{sec:naive_result}.

\subsubsection{Random forest classifier}

Random forest~\citep{breiman2001random} is an ensemble learning method that combines decision trees to perform a classification exercise.
It utilizes the bagging technique to create different subsets of the original dataset by sampling with replacement.
In addition to bagging, random forest incorporates randomness into the feature selection process.
The above two aspects ensure that each decision tree in the random forest model is built on different subsets of samples and features, which reduces overfitting for high-dimensional data.
Random forest determines the final output by combining the predictions of all decision trees through majority voting.
If the majority of trees predict that a firm will achieve high growth in the future, then it will be classified as an HGF.
We construct the random forest classifier utilizing the Python implementation provided by the scikit-learn library\footnote{\url{https://github.com/scikit-learn/scikit-learn}}.
Table~\ref{tab:5} lists the considered hyper-parameters of our random forest classifiers.

\subsubsection{XGBoost classifier}

XGBoost~\citep{chen2016xgboost} is an ensemble learning method that leverages the boosting technique, where decision trees are built in a sequential manner.
During the training process, the algorithm addresses bias by assigning higher weights to the misclassified samples from the previous trees, thereby emphasizing these challenging samples in subsequent training steps.
For the final prediction, XGBoost employs a weighted averaging process that assigns different weights to all trees based on their predictive power.
We construct the XGBoost classifier utilizing the Python implementation provided by the XGBoost library\footnote{\url{https://github.com/dmlc/xgboost}}.
Table~\ref{tab:5} lists the considered hyper-parameters of XGBoost classifiers.

\begin{table}[pos=hbt]
\centering
\caption{Parameter set of predictive models.}
\resizebox{\linewidth}{!}{
\begin{tabular}{p{3cm}p{5cm}p{9cm}}
\hline
Classifier & Parameter & Search space \\ 
\hline
\multirow{3}{*}{Random forest} & Number of decision trees & \{100, 200, 500, 1000\} \\
                               & Maximum tree depth & \{3, 5, 10, 20, 30, No limit (None)\} \\
                               & Weights for different classes & \{Weights inversely proportional to class frequencies (Balanced),  \newline
                               Equal weights for all classes (None)\} \\
\hline
\multirow{5}{*}{XGBoost}       & Number of decision trees & \{100, 200, 500, 1000\} \\
                               & Maximum tree depth & \{3, 5, 10, 20, 30, No limit (None)\} \\
                               & Minimum loss reduction & \{0.0, 0.1, 0.2, 0.3, 0.4, 0.5\} \\
                               & Learning rate & \{0.0, 0.1, 0.2, 0.3, 0.4, 0.5\} \\
                               & Weights for different classes & \{Weights inversely proportional to class frequencies (Balanced), \newline
                               Equal weights for all classes (None)\} \\
\hline
\end{tabular}}
\label{tab:5}
\end{table}

\subsubsection{Cross-validation}

Given the limited number of firms in our dataset, we randomly shuffle the dataset $100$ times to mitigate the potential bias introduced by the specific order of the firms.
For each shuffle, we conduct a nested cross-validation (CV) to obtain an unbiased estimate of the model's generalization performance by accounting for the variability introduced by the hyper-parameter optimization process.
The nested CV involves two loops: the outer CV loop for evaluating model performance and the inner CV loop for tuning hyper-parameters.
The detailed steps of a nested CV are as follows:
\begin{enumerate}
    \item {Conduct a $3$-fold stratified CV to split the dataset into $3$ folds for the outer CV loop. The stratified sampling ensures that the proportion of HGFs is identical in each fold.}
    \item {For each outer fold:
        \begin{enumerate}
        \item {Hold out that fold as the testing set and use the remaining $2$ folds as the training set.}
        \item {Conduct another $3$-fold stratified CV to split this training set for the inner CV loop. For each fold in the inner CV loop, hold out one fold as the validation set and use the remaining $2$ folds as the training subset.}
        \item {Train the model with different hyper-parameters on the training subset and select the best hyper-parameters based on the performance on the validation set. The hyper-parameters are tuned using randomized search~\citep{bergstra2012random}.}
        \item {Train the model with the best hyper-parameters on the entire training set.}
        \item {Compute the model performance on the testing set.}
        \end{enumerate}
    }
    \item {Repeat for all outer folds and compute the average model performance.}
\end{enumerate}

\subsection{Evaluation}

The objective of our predictive experiments is to forecast firms that will achieve high economic growth in the later period, utilizing their information on financial performance and patenting activities in the early period.
To evaluate the performance of predictive models, we employ Precision and the area under the Precision-Recall curve (AUPRC).
Precision measures the proportion of correctly predicted positive samples (i.e., HGFs) out of the total predicted positive samples, which specifically assesses the effectiveness of models in identifying the positive class.
Recall measures the proportion of correctly predicted positive samples out of the total actual positive samples, which essentially evaluates the model's ability to identify all positive samples.
In our dataset, the negative class (account for 90\%) significantly outnumbers the positive class (account for 10\%).
This class imbalance can mislead both Precision and Recall.
Besides, the value of Precision and Recall strongly depends on the probability threshold, which determines whether a sample with a given probability score is classified as positive or negative.

To achieve a more comprehensive and robust evaluation, we select AUPRC as the other evaluation metric.
The Precision-Recall curve illustrates the values of Precision and Recall of the model at different probability thresholds.
AUPRC considers the trade-off between Precision and Recall across these various probability thresholds, which makes it less sensitive to class imbalance and thus appropriate for evaluating models on imbalanced datasets~\citep{saito2015precision}.

\section{Results}

\subsection{Model-free evidence}\label{sec:naive_result}

\begin{figure*}[pos=hbt]
    \centering
    \caption{Predictive performance of the naïve classifiers under different growth indicators.
    Each bar represents an individual feature: red bars for financial features, blue bars for technological features, and green bars for network-based features.
    The error bars indicate the standard error of the classifiers' average performance, calculated over $100$ different shuffles of the dataset.
    The inset shows the top $7$ features with the best performance. Financial features have the strongest predictive performance; technological features, such as the maximum economic value of a firm's granted patents, are strongly associated with high economic growth.}
    \includegraphics[width=0.95\linewidth]{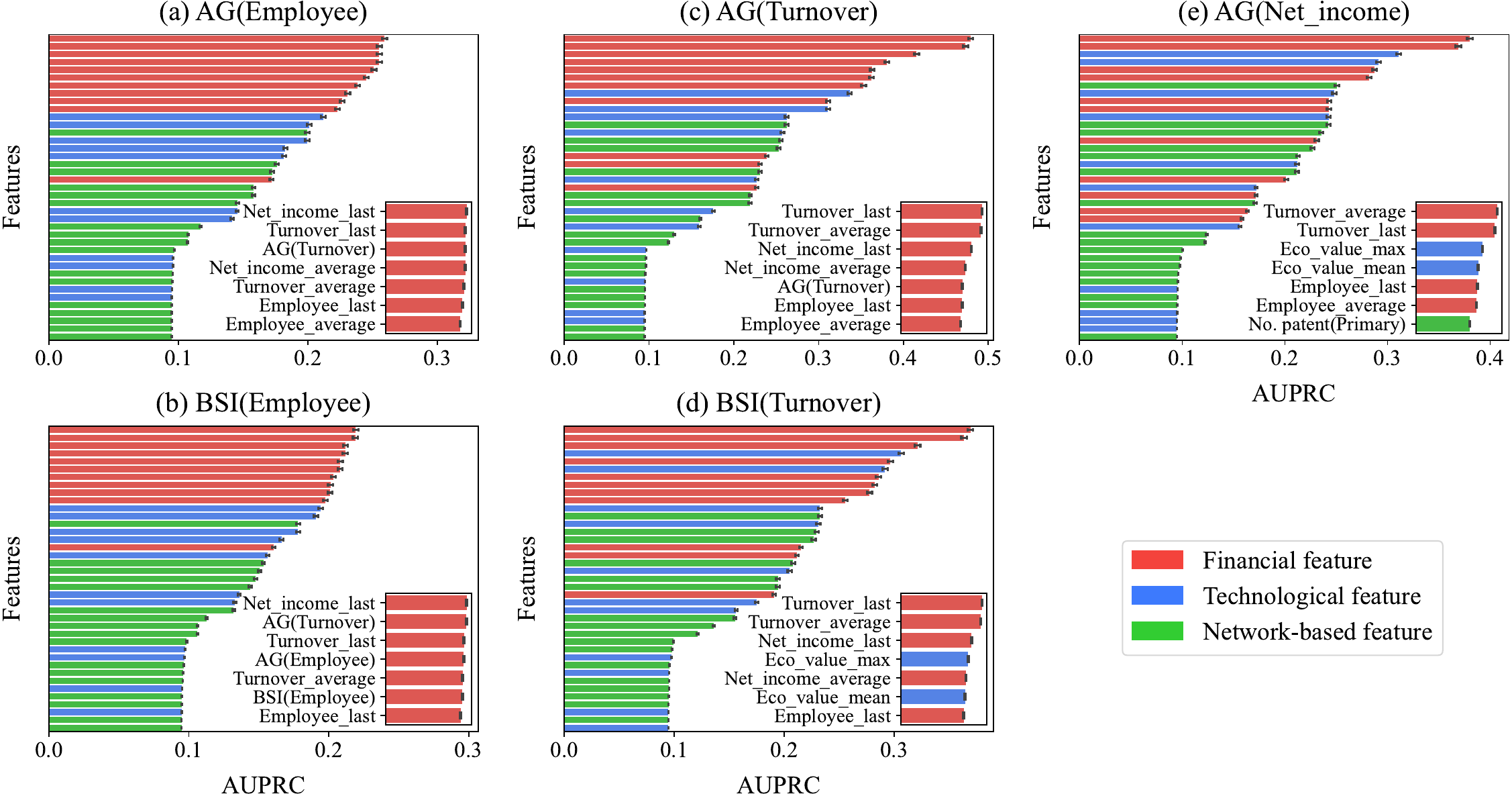}
    \label{fig:3}
\end{figure*}

To gain initial insights into the predictive importance of features, we begin by evaluating the performance of naïve classifiers in predicting HGFs by using individual features.
The results are illustrated in Figure~\ref{fig:3}, where we sort the features by predictive performance. The colors indicate the feature category and we highlight the top-performing features in the inset.
Overall, financial features exhibit the strongest predictive power, especially in relation to employee number growth; for the growth of turnover and net income, the classifiers utilizing patents' economic value demonstrate comparable performance to those relying on financial features, which highlights the predictive potential of technological features.
The predictive power of both technological and network-based features is highly heterogeneous.
While most features perform at a similarly level, a few features exhibit significantly higher predictive power in comparison.
Among the technological feature-based classifiers, the best-performing one is based on \textit{Eco\_value\_max}, which indicates a notable correlation between the maximum economic value of a firm's granted patents and high economic growth.
For example, the classifier using \textit{Eco\_value\_max} achieves an AUPRC of $0.31$ in Fig.~\ref{fig:3} (e), which outperforms most financial feature-based classifiers.
Among the network-based feature-based classifiers, the one utilizing \textit{No. patent (Primary)} exhibits the best predictive performance.
This suggests that high economic growth is strongly correlated with the frequency of patenting activities in primary technologies.
For example, the classifier using \textit{No. patent (Primary)} exhibits performance that surpasses most classifiers in Fig.~\ref{fig:3} (e), though its performance is still lower than the classifier using \textit{Eco\_value\_max}.
In the following, we show how a more refined analysis with the machine learning models extends these preliminary insights.

\subsection{Models' predictive performance}

To assess the impact of introducing technological and network-based features on HGF prediction, we conduct a comparative analysis by comparing the performance of models that utilize different feature sets as inputs.
Specifically, we set the baseline model, referred to as \textit{F}, as the one only utilizing financial features.
Then we consider the following comparative models: model \textit{F+T} that incorporates both financial and technological features, model \textit{F+N} that incorporates financial and network-based features, and model \textit{F+T+N} that incorporates all available features.
The average improvement rate of these comparative models against the baseline model is calculated over $100$ iterations of $3$-fold cross-validations, as presented in Table~\ref{tab:6}.
The highest values of each metric-growth indicator couple are in bold.

First of all, we note that, among the improvement rates for the three comparative models across five growth indicators, nearly all improvements are positive.
This consistency proves that both technological and network-based features bring valuable information for HGF prediction, leading to better-performing models with respect to the ones that use financial features only.
Our first main finding can, therefore, be stated as follows:
\begin{mf}
Incorporating financial features with either technological, network-based features, or both, leads to more accurate HGF predictions compared to using financial features alone.
\end{mf}

To further evaluate the relative importance of technological versus network-based features, we compare the performance of model \textit{F+T} and \textit{F+N}.
The results reveal that which model among \textit{F+T} and \textit{F+N} outperforms depends on the considered target variable, which indicates that both technological features and network-based features are essential in HGF prediction.
Furthermore, model \textit{F+T+N} exhibits the best performance in most cases.
Its superiority compared to model \textit{F+T} and \textit{F+N} reveals that the combination of technological and network-based features helps improve predictive performance.

\begin{mf}
The best-performing model includes both technological and network-based features for almost all the target variables considered here, which indicates that both classes of features play a crucial role in HGF prediction.
\end{mf}

In summary, the analysis of models' predictive performance proves the effectiveness of incorporating both technological and network-based features with financial features for HGF prediction.
The strong predictive power of the combined model motivates us to perform a detailed feature importance analysis based on the \textit{F+T+N} model, which we shall present next.

\begin{table}[pos=hbt]
\centering
\caption{Predictive performance of random forest models, including the baseline model's original performance (model $F$, which uses Financial features only) and the improvement rates of the comparative models (which incorporate Technological and Network-based features). Exploiting patenting activities and economic complexity methods vastly enhances the predictive performance.
The highest values are in bold. All values are averaged over $100$ iterations of $3$-fold cross-validations.}
\resizebox{\linewidth}{!}{
\begin{tabular}{p{2cm}p{3cm}p{3cm}p{3cm}p{3cm}p{3cm}}
\hline
\multirow{2}{*}{Metric} & \multirow{2}{*}{Growth indicator} & \multicolumn{4}{l}{Input features}    \\ \cline{3-6} 
& & F & F+T & F+N & F+T+N \\
\hline
\multirow{5}{*}{Precision}
& AG (Employee)  & 0.335 ($\pm$0.011) & \textbf{+72.46\%} ($\pm$3.6\%) & +44.29\% ($\pm$3.3\%) & +68.41\% ($\pm$3.6\%) \\
& BSI (Employee)  & 0.344 ($\pm$0.015) & +77.24\% ($\pm$5.7\%) & +63.51\% ($\pm$6.0\%) & \textbf{+79.37\%} ($\pm$5.9\%) \\
& AG (Turnover)  & 0.578 ($\pm$0.007) & +14.93\% ($\pm$1.8\%) & +18.95\% ($\pm$1.6\%) & \textbf{+20.31\%} ($\pm$1.8\%) \\
& BSI (Turnover)  & 0.637 ($\pm$0.008) & +15.85\% ($\pm$1.9\%) & +10.81\% ($\pm$1.7\%) & \textbf{+17.57\%} ($\pm$1.8\%) \\
& AG (Net\_income)  & 0.673 ($\pm$0.004) & +0.28\% ($\pm$0.7\%) & \textbf{+1.94\%} ($\pm$0.8\%) & -0.33\% ($\pm$0.7\%) \\
\hline
\multirow{5}{*}{AUPRC}
& AG (Employee)  & 0.38 ($\pm$0.002) & \textbf{+24.22\%} ($\pm$0.7\%) & +14.74\% ($\pm$0.6\%) & +22.96\% ($\pm$0.7\%) \\
& BSI (Employee)  & 0.32 ($\pm$0.002) & +24.61\% ($\pm$0.8\%) & +15.75\% ($\pm$0.7\%) & \textbf{+25.32\%} ($\pm$0.8\%) \\
& AG (Turnover)  & 0.553 ($\pm$0.003) & +16.38\% ($\pm$0.5\%) & +14.37\% ($\pm$0.5\%) & \textbf{+17.03\%} ($\pm$0.5\%) \\
& BSI (Turnover)  & 0.488 ($\pm$0.003) & +19.98\% ($\pm$0.5\%) & +14.15\% ($\pm$0.5\%) & \textbf{+20.13\%} ($\pm$0.6\%) \\
& AG (Net\_income)  & 0.564 ($\pm$0.003) & +4.71\% ($\pm$0.4\%) & +5.65\% ($\pm$0.4\%) & \textbf{+6.65\%} ($\pm$0.4\%) \\
\hline
\end{tabular}}
\label{tab:6}
\end{table}

\subsection{Feature importance}\label{sec:feature_importance}

To measure the contribution of each feature in the random forest models, we compute the Gini importance score as the total reduction of the Gini impurity brought by a feature across all trees in the forest~\citep{breiman1984classification}.
The Gini impurity is a metric that quantifies the probability of incorrectly classifying a randomly chosen sample.
A detailed introduction of the Gini importance and Gini impurity is provided in Appendix~\ref{sec:appendix_gini}.
Figure~\ref{fig:4} shows the rankings of the top $12$ feature with the highest Gini importance score in model \textit{F+T+N}, where the inset shows the rankings of all features.

As one may expect, the relative contribution of features depends on the specific growth indicators.
For the indicators related to the number of employees, financial features demonstrate the dominant place in predictive importance.
But for the indicators related to turnover and net income, some technological features outperform most financial features, which is consistent with the results of naïve classifiers discussed in Section~\ref{sec:naive_result}. The rising places of technological features indicate their advantages in predicting firms with high growth of profitability.

When looking at the relative feature importance within each class of features, remarkable regularities are observed.
As for financial features, those related to the corresponding target growth indicators rank higher, as one would expect.
For example, in Figs.~\ref{fig:4} (a)--(b), features \textit{Employee\_last} and \textit{Employee\_average} take the top places.

Among technological features, the economic values of firms' patents have more predictive power than the technological values of firms' patents for all target variables.
Specifically, the maximum economic value of a firm's granted patents has higher importance score than the technological values (measured based on citation counts, see~\ref{sec:tech_feature}).
This result extends the findings in~\cite{xu2024citations} from predicting the value of a firm's future patents to the economic growth prediction studied here.
We can summarize this finding as follows:

\begin{mf}
Among the technological features, the maximum economic value of a firm's granted patents tends to be the most important predictor of growth. In particular, it is more important than citation-based indicators for all the analyzed target variables.
\end{mf}

The higher importance of patents' economic value confirms that a firm's potential to derive economic gains from technological innovation is a key factor in achieving high economic growth.
As for network-based features, we observe that the number of patents related to a firm's primary technologies consistently holds the highest importance:
\begin{mf}
Among the network-based features, the number of patents related to a firm's primary technologies consistently holds the highest importance.
\end{mf}
This finding qualitatively aligns with the study carried by ~\cite{pugliese2019coherent}, which suggested that a combination of the coherence and diversification of firms' technological portfolios is positively correlated with their financial performance.
Most of the analyzed network-based features tend to exhibit low importance scores.
This result suggests that the predictive potential of network-based features has yet to be fully exploited by the current methods employed in network analysis.
Future studies may generalize existing network-based features---such as the Fitness-Complexity scores---to capture the specific characteristics of the firm's economic and technological ecosystems~\citep{laudati2023different}.

\begin{figure*}[pos=hbt]
    \centering
    \caption{Rankings of the top $12$ features with the highest Gini importance scores under different growth indicators.
    Each bar corresponds to a feature.
    The dashed lines represent the average importance scores across all the features considered.
    The error bars indicate the standard error of the features' average importance scores, calculated over $100$ iterations of $3$-fold cross-validations.
    The inset shows the importance rankings of all features. While the growth of the number of employees is associated with financial features, the maximum economic value of a firm's granted patents is a key feature to predict turnover and net income. Among the network-based features, the most important one is the number of patents related to a firm's primary technologies.}
    \includegraphics[width=0.95\linewidth]{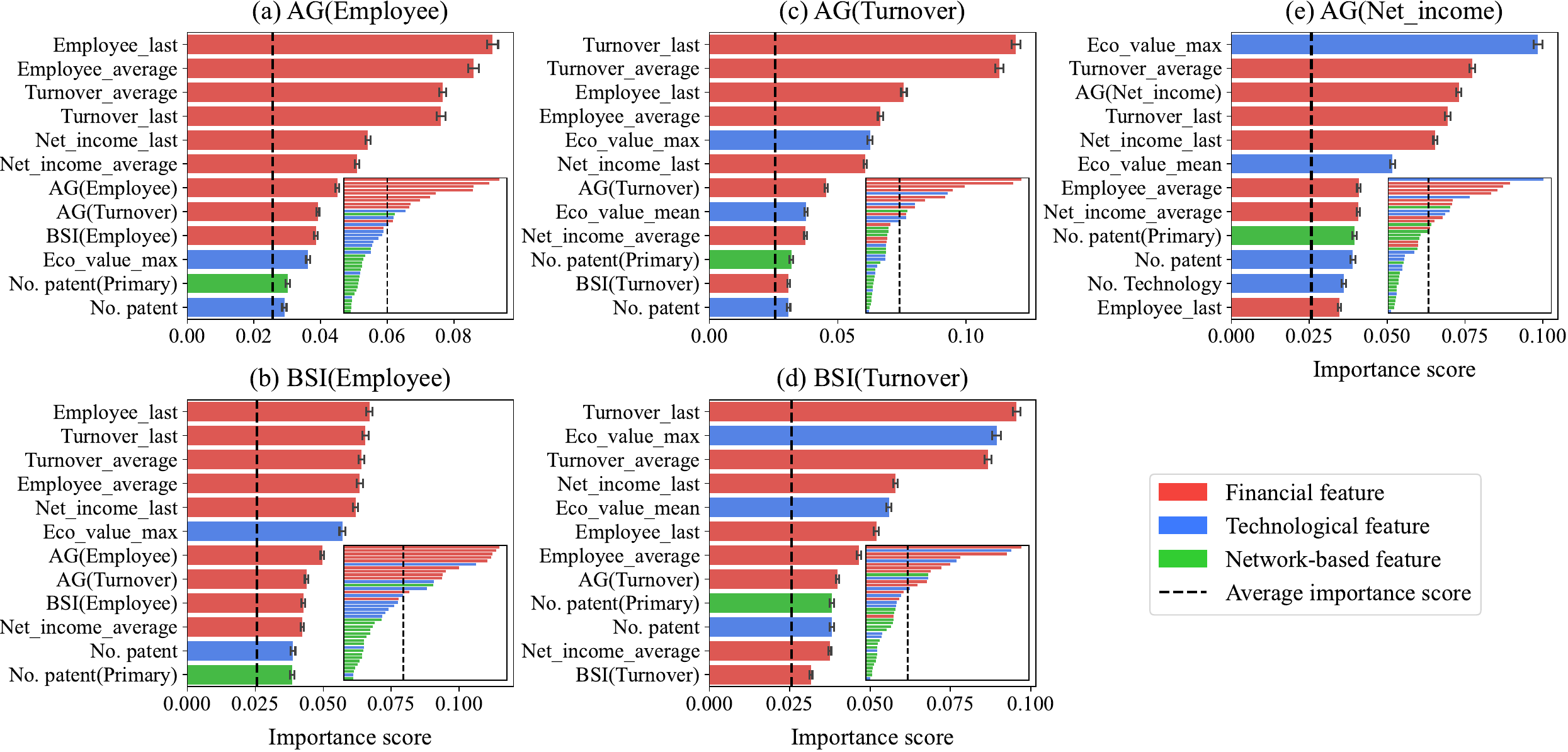}
    \label{fig:4}
\end{figure*}

The Gini importance analysis illustrated in Fig.~\ref{fig:4} assesses the general influence of features in building a predictive model. However, the importance of different features may vary from firm to firm. To identify the specific contribution of each feature in the prediction of individual firms, we resort to the SHapley Additive exPlanations (SHAP)\footnote{\url{https://github.com/shap/shap}}~\citep{lundberg2017unified,lundberg2020local2global}, an interpretable framework rooted in cooperative game theory.
Within the SHAP framework, the feature importance score of each feature is computed for each firm, which represents the feature's contribution to the final prediction for that firm.
A positive (negative) SHAP value signifies that the feature value positively (negatively) contributes to increasing the probability of achieving high economic growth for that firm.
The absolute SHAP value reflects the magnitude of a feature's contribution.
Moreover, we can also measure the importance of a feature as its average absolute SHAP value over all firms, as shown in Appendix~\ref{sec:appendix_shap_importance}.
Compared to Fig.~\ref{fig:D.1}, the feature importance rankings measured by the two methods are robust.

\begin{figure*}[pos=hbt]
    \centering
    \caption{The SHAP values of the top $12$ features with the highest SHAP-based importance scores.
    There is a positive correlation between the feature values and SHAP values for most features, which indicates that firms with higher values in these features have a higher probability of achieving high economic growth.}
    \includegraphics[width=0.95\linewidth]{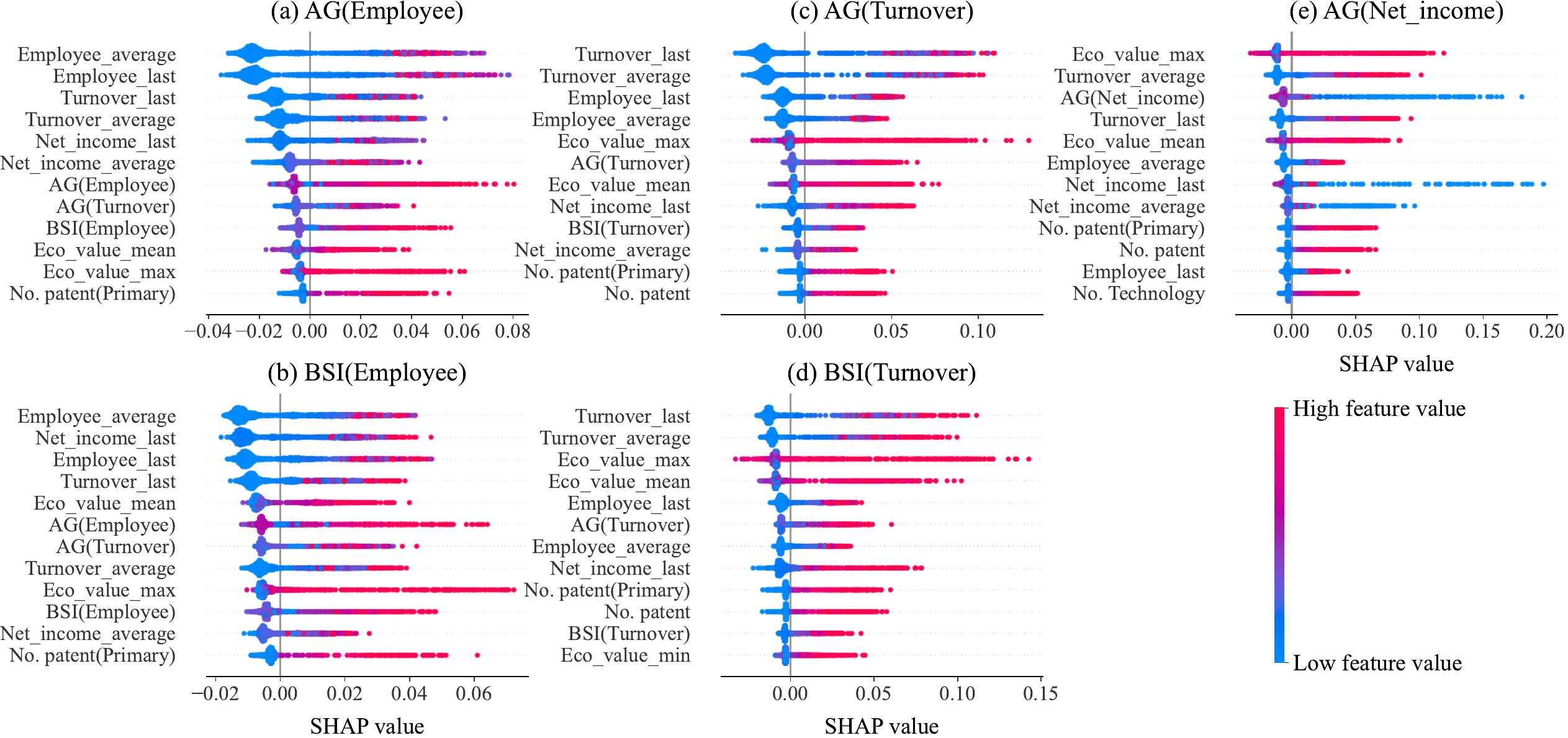}
    \label{fig:5}
\end{figure*}

Figure~\ref{fig:5} shows the SHAP value of the top $12$ features with the highest SHAP-based importance scores, in terms of model \textit{F+T+N} under different growth indicators.
Each row represents a feature, while each point corresponds to a firm.
The color of a point reflects the feature value, with red indicating a high value and blue indicating a low value.
The horizontal position of a point indicates the SHAP value of the feature for that firm, which is averaged over $100$ iterations of $3$-fold cross-validations.

We find that the values of most features are positively correlated with their corresponding SHAP values, which indicates that firms with higher values in these features have a higher probability of achieving high economic growth.
However, some exceptions still exist.
For instance, in Figs.~\ref{fig:5} (c)--(e), some high values of \textit{Eco\_value\_max} and \textit{Eco\_value\_mean} correspond to negative SHAP values.
This means that patents with high economic potential are less likely to show a positive impact on achieving industry-leading growth for some firms.
In Fig.~\ref{fig:5} (e), there is a negative correlation between the feature values and SHAP values for feature \textit{AG (Net\_income)}.

In the supplementary material (see Appendix~\ref{appendix:sm}), the results within different time windows are summarized in Section~S1, and the results based on XGBoost models are presented in Section~S2.
Our analysis demonstrates the robustness of our findings across different time windows and predictive models.

To summarize, the feature importance rankings emphasize the crucial role of financial features in HGF prediction, followed by technological features, and network-based features ranking last.
Technological features, especially for the economic values of a firm's granted patents, occupy the most important positions in predicting firms with high growth in profitability.
The minimal predictive importance of network-based features calls for more effective network analysis methods to extract meaningful firm features.
Based on the SHAP framework, we can evaluate the magnitude and direction of each feature's impact on individual firms.
Our analysis reveals that for most features, a higher value positively correlates with a firm's likelihood of becoming an HGF.
However, we also observe some exceptions to this general trend, which will be further analyzed in the next subsection.

\subsection{Partial dependence results}

To explore the relationship between features and the predicted outcome, we draw the partial dependence plots (PDP) to visualize the marginal effects of features on the model output~\citep{friedman2001greedy}.
PDPs are visualization tools for interpreting the dependence of a model's predicted output on a subset of the input features.
To create a PDP, a set of samples is generated by varying the values of the selected feature while holding all other features constant.
These samples are then input into the trained predictive model to calculate the average predicted outcome for each feature value.
This process reveals the partial dependence of the predicted outcome on the selected feature.
There are two commonly used forms of PDPs~\citep{molnar2022}: one-way PDPs focus on illustrating the effect of a single feature on the predicted outcome, while two-way PDPs demonstrate the interactive effect of two features.
To explain the exceptions in Fig.~\ref{fig:5}, we focus on two features with high importance scores: \textit{Employee\_last} representing the size of firms, and \textit{Eco\_value\_max} representing the quality of firms' innovative activities.
All PDPs are built using the scikit-learn library.

\begin{figure*}[pos=hbt]
    \centering
    \caption{The one-way PDPs of \textit{Employee\_last} and \textit{Eco\_value\_max} under specific growth indicators.
    (a) Small and medium-sized firms are positively affected by their size since the probability of being an HGF increases with the number of employees. However, this influence is absent once about $15,000$ employees are reached.
    (b) The maximum economic value of a firm's granted patents is positively associated with the probability of high growth, but only if that economic value exceeds a threshold of approximately $0.7$.}
    \includegraphics[width=0.95\linewidth]{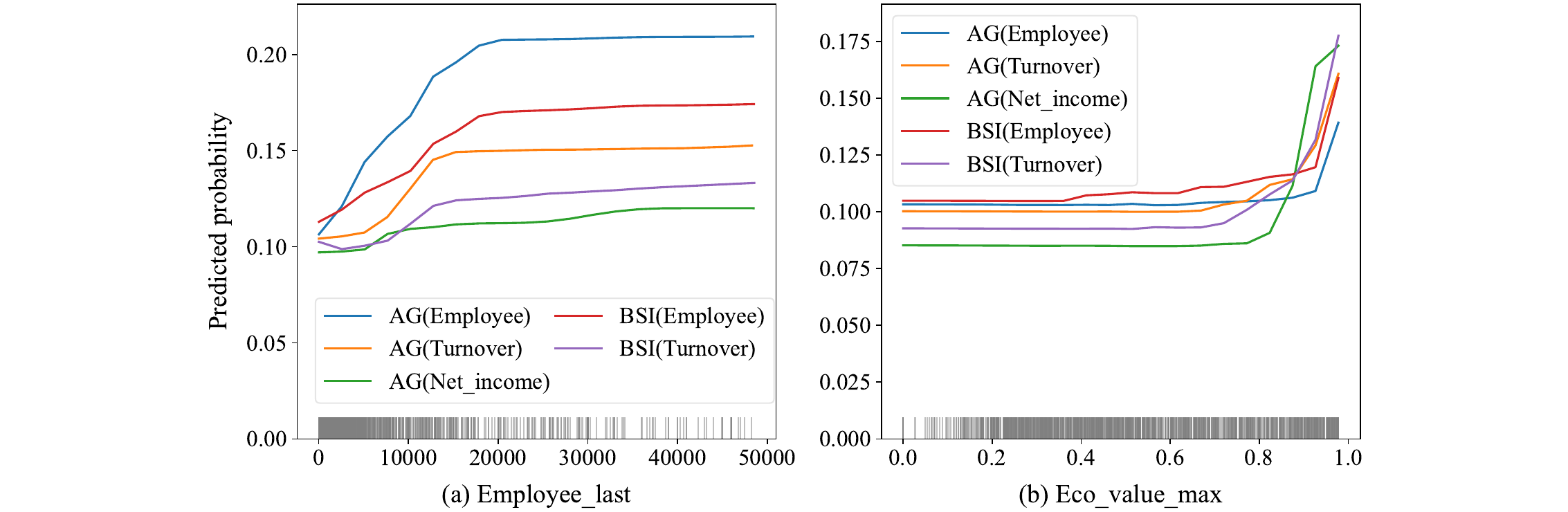}
    \label{fig:6}
\end{figure*}

Figure~\ref{fig:6} shows how changes in feature values affect predicted results, with the horizontal axis representing the feature values and the vertical axis indicating the predicted probability of high growth under specific growth indicators.
Both plots indicate that the relationship between a single feature and the probability of high growth is nonlinear.
Specifically, in Fig.~\ref{fig:6} (a), the probability of high growth keeps increasing with the number of employees up to approximately $15,000$, after which the growth significantly slows down, and finally holding steady.
This can be summarized as follows:
\begin{mf}
Firm size is positively associated with high-growth probability up to a certain threshold size, after which the association plateaus.
\end{mf}
First, the initial phase of rapid growth in predicted probability corresponds to the trend of small firms rapidly expanding to reach the so-called minimum efficient scale~\citep{oliveira2006testing,moreno2007high}.
At this scale, firms can produce at the lowest possible cost per unit, increasing their likelihood of survival and competitiveness in the marketplace~\citep{arbelo2022smes}.
Then, the plateau in the predicted probability may reflect the risks faced by large firms.
Although firm size growth facilitates resource accumulation, it simultaneously increases the adjustment costs of transforming these resources into future productive opportunities~\citep{lockett2013organic,zhou2019there}, such as the time and effort needed to integrate new employees into the firm~\citep{penrose1995theory}.
Consequently, excessive size growth may cause firms to reach an inflection point where the adjustment costs overtake the benefits of high growth, even going to potentially endangering a firm's survival~\citep{pierce2013too}.

However, the relationship between high-growth probability and the maximum economic value of a firm's granted patents is the opposite.
In Fig.~\ref{fig:6} (b), the probability of high growth begins to increase only after the feature value reaches a threshold value (approximately $0.7$). We can summarize this as follows:
\begin{mf}
The maximum economic value of a firm's granted patents is positively linked to high-growth probability only after a threshold value is exceeded.
\end{mf}
The economic value of a patent reflects the stock market reactions to the patent's grant.
A patent with high economic value signifies market recognition of the potential to derive economic benefits from transforming it into commercial products or services.
This recognition amplifies the likelihood of firms achieving high economic growth through the successful transformation of high-value patents.
Conversely, for firms with low \textit{Eco\_value\_max}, all of their patents lack perceived potential, which might fail to increase the probability of attaining significant economic benefits through patent transformation.
This could potentially explain the previous main finding.

\begin{figure*}[pos=hbt]
    \centering
    \caption{The two-way PDPs between \textit{Employee\_last} and \textit{Eco\_value\_max} under specific growth indicators.
    The probability of being an HGF in terms of the number of employees (panels a--b) is more dependent on having a patent with high economic value than on having a large size.
    On the contrary, regarding turnover and net income (panels c--e), the HGF probability is less dependent on this measure of patenting activities.
    For firms with over approximately $40,000$ employees, firm size shows a stronger impact.
    }
    \includegraphics[width=0.95\linewidth]{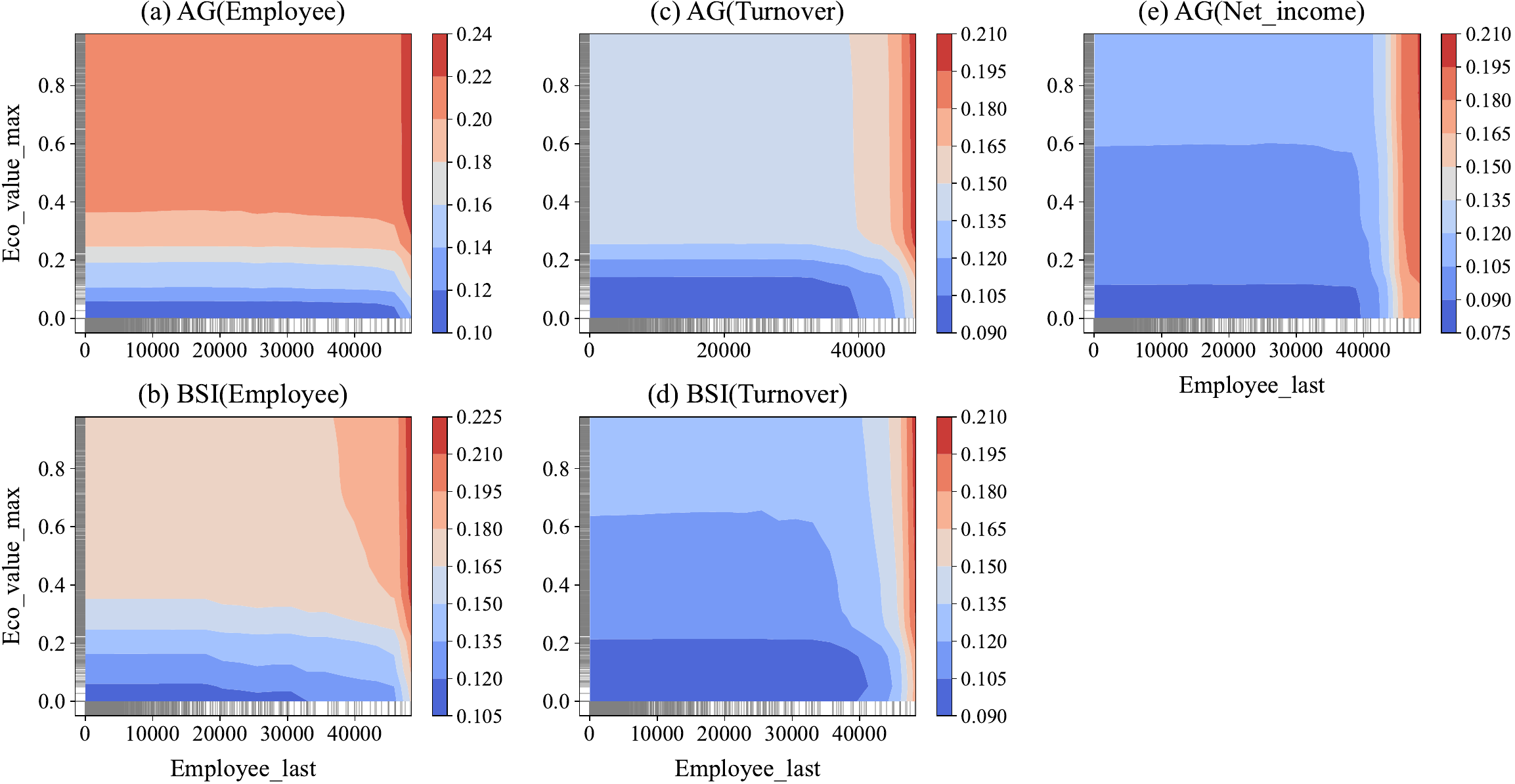}
    \label{fig:7}
\end{figure*}

Figure~\ref{fig:7} visualizes the interactive effect of \textit{Employee\_last} (horizontal axis) and \textit{Eco\_value\_max} (vertical axis) on the predicted probability (see the legend).
In Figs.~\ref{fig:7} (a)--(b), \textit{Eco\_value\_max} shows a greater impact on the predicted results of firm size growth.
Specifically, the predicted probability significantly increases as \textit{Eco\_value\_max} increases along the vertical axis.
In Figs.~\ref{fig:7} (c)--(e), we find a clear boundary in firms' size located at approximately $40,000$ employees.
For firms with less than $40,000$ employees, \textit{Eco\_value\_max} still holds the greater influence, but the maximum predicted probability is relatively low.
When the size exceeds the boundary, the maximum predicted probability increases significantly, and \textit{Employee\_last} shows a stronger impact on the horizontal axis.
The changing impact of \textit{Eco\_value\_max} under different firm sizes may explain the exception in Fig.~\ref{fig:5} that the high values of \textit{Eco\_value\_max} in many firms correspond to negative SHAP values, in term of the growth of turnover and net income.
Large firms may have superior commercialization capabilities to extract value from patents~\citep{arora2023invention}, while for small and medium-sized firms, their current capacity often falls short of fully exploiting the economic benefits of high-value patents without undergoing an expansion process.
Consequently, high-value patents are more likely to positively impact larger firms in achieving industry-leading growth rather than small and medium-sized firms, primarily due to the substantial costs associated with R\&D and increased employment.

\begin{figure*}[pos=hbt]
    \centering
    \caption{The proportion of positive and negative SHAP values of feature \textit{Eco\_value\_max} in different firm groups.
    In the groups with larger firms, the proportion of positive SHAP values is higher than that of negative SHAP values.
     This reveals that high-value patents are more likely to have a positive impact on larger firms, contributing to industry-leading growth in turnover and net income.}
    \includegraphics[width=0.95\linewidth]{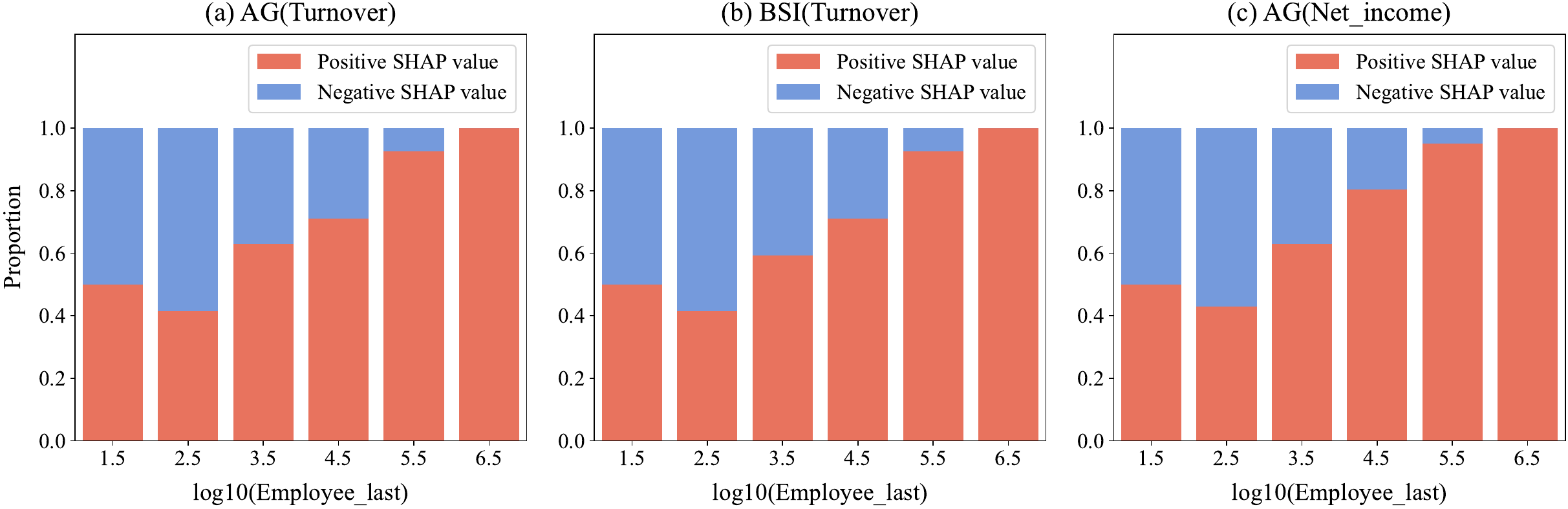}
    \label{fig:8}
\end{figure*}

To further validate the varying impacts of high-value patents on firms of different sizes, we focus on firms with high-value patents ($Eco\_value\_max>0.7$) and group them based on their size (measured by \textit{Employee\_last}).
Fig.~\ref{fig:8} illustrates the proportion of positive and negative SHAP values of feature \textit{Eco\_value\_max} across different firm groups.
In the groups with larger firms, positive SHAP values account for a higher proportion than negative SHAP values, which validates the above analysis that high-value patents are more likely to have a positive impact on larger firms in achieving industry-leading growth in turnover and net income.
Thus, we can summarize:
\begin{mf}
Patents with high economic value are more likely to positively impact larger firms in achieving industry-leading growth in turnover and net income, compared to small and medium-sized firms.
\end{mf}

With the help of PDP, we analyze the individual and interactive effects of the two most significant features, \textit{Employee\_last} and \textit{Eco\_value\_max}, on the predicted probability of a firm's high future growth.
The individual effects are nonlinear.
Overall, larger firms have a higher probability of high growth below a threshold size, while smaller firms can obtain a greater increase in growth potential through size expansion.
Furthermore, only patents with high economic values can significantly increase the predicted probability of high growth.
As for the interactive effect, we reveal the different impacts of high-value patents on smaller and larger firms.
High-value patents typically contribute positively to the high growth of profitability for larger firms, due to their effective ability to transform technological innovations into economic benefits.
In contrast, smaller firms have a lower probability of achieving industry-leading growth from high-value patents, potentially because they need substantial additional investments to drive similar commercial transformations.

\section{Discussion}

This work provides an extensive comparison of the predictive power of diverse financial, technological, and network-based features for the forecast of high-growth firms.
This is possible thanks to a dataset that incorporates the information of firms' financial statements, patenting activities, and network-based measures.
The predictive performance comparison reveals the effectiveness of incorporating technological innovation activities coupled with network analysis methods into the evaluation of firms' future financial performance.
With the application of explainable artificial intelligence methods, we are able to quantify the prominent importance of patents' economic value in predicting high-growth firms.
This finding emphasizes the necessity to analyze firms' technological capabilities based on the inherent attributes of their patenting activities beyond macro-level statistics.
The plateau effect of large firm size and low patents' economic value on the predicted probability of high growth reflects the nonlinear relationship between these single features and the results.
In summary, explainable machine learning methods provide a clear view of the influence of single corporate characteristics on achieving high economic growth; this offers a solid theoretical basis for managers' decisions.

Our empirical results clearly indicate that, to predict high-growth firms, the most effective model integrates technological and network-based features with the baseline financial features.
This model can be used by managers for evaluating the future financial performance of firms, monitoring the health of a given firm's state beyond its financials, and informing resource allocation decisions.
By comparing the predicted probability of high growth when adjusting the values of certain features, managers can make preliminary estimates of the future effects of their decisions.
This practical application helps in strategic planning and operational adjustments.
The analysis based on explainable machine learning methods reveals the average predictive effects of features among all firms, which provides macro-level insights for firm growth.
For instance, the predicted high-growth probability increases with firm size up to a certain threshold, which reflects the positive impacts of size expansion for small firms.
After this threshold, the predicted probability plateaus despite further increases in firm size.
This plateau effect indicates limited improvements from size expansion for large firms.
The varying impacts of size expansion on firms of different sizes suggest that managers should adopt a cautious attitude toward scaling and develop strategies based on the firm's current state.

This work can be extended in a number of directions by future research efforts.
First, here we only consider firms listed in the stock market, which excludes startups and firms that fail early on in their lifecycle.
Including firms that fail early on is a promising direction for future research, yet it would likely to require the collection of new data compared to those analyzed here.
Second, among the three classes of features studied here, many network-based features hold low importance in predictions, and the improvements in predictive performance brought by the most important network-based features is lower than those brought by the most important technological features.
Hence, a promising direction for future studies is to design new network-based methods to infer firms' competitiveness from network structures, by factoring in the specific structural features of the networks firms are embedded in.
If successful, these efforts could highlight the importance of network-based features, thereby further improving the performance in HGF prediction and tightening the link between the economic complexity field and HGF prediction studies.
Third, we stress that all the results presented here are predictive, and not causal. Revealing the causal mechanisms behind our findings is a key direction for future research, as it could both validate some of our conjectures (e.g., on the reasons behind the partial dependence plot results) and provide solid implications for managerial decision-making.

% \printcredits

\section*{Declaration of competing interest}

The authors declare that they have no known competing financial interests or personal relationships that could have appeared to influence the work reported in this paper.

\section*{Acknowledgement}

LL acknowledges the National Natural Science Foundation of China (Grant No. T2293771), and the New Cornerstone Science Foundation through the XPLORER PRIZE. 
SX acknowledges the National Natural Science Foundation of China (Grant No. 62306191).
MSM acknowledges financial support from the URPP Social Networks and the Swiss National Science Foundation (Grant No. 100013--207888).
AZ acknowledges the PRIN project No. 20223W2JKJ “WECARE”, CUP B53D23003880006, financed by the Italian Ministry of University and Research (MUR), Piano Nazionale Di Ripresa e Resilienza (PNRR), Missione 4 “Istruzione e Ricerca” - Componente C2 Investimento 1.1, funded by the European Union - NextGenerationEU.

\section*{Data availability}
The authors do not have permission to share data.
The patent data used in this study are openly available at \url{https://www.mikewoeppel.com/data}.
The financial data used in this study can be accessed from the Orbis database (\url{https://www.bvdinfo.com}), which may require institutional access or a subscription.

\FloatBarrier
\newpage
\section*{Appendix}
\appendix

\section{Details of data matching process}\label{sec:appendix_matching}
In Section~\ref{sec:collection}, for each firm identifier (permco) in the USPTO dataset, we retrieve its corresponding firm name and ticker symbol from the WRDS platform.
Then we input these ticker symbols into the Orbis database to retrieve the corresponding financial data.
However, ticker symbols can be reused by different firms.
To ensure that the financial data we gather from Orbis is accurately matched with the correct firms from WRDS, we implement a matching process to compare the firm names.
The detailed matching steps are as follows:
\begin{enumerate}
    \item {Standardize firm names. Replace the common terms in firm names with their abbreviated forms. For example, replace ``corporation'' with ``corp'' and ``holdings'' with ``hldgs''.}
    \item {Remove conflicting identifiers. The combination of firm name and ticker symbol is used as the identifier for matching. Remove the data with identifiers that correspond to multiple firms to maintain the uniqueness of identifiers in both Orbis and WRDS datasets.}
    \item {Compute matching scores. Use the fuzzy matching package\footnote{\url{https://github.com/seatgeek/fuzzywuzzy}}~\citep{cohen2011fuzzywuzzy} to calculate the Levenshtein distance between the firm names that share the same ticker symbol. If the matching score reaches $80$ (out of $100$), the matching is considered to be successful.}
    \item {Merge the data. After successfully matching the data from Orbis and WRDS, merge the patent data and financial data, further leading to the final dataset that includes complete information for each firm.}
\end{enumerate}

\section{Details of network-based features}

\subsection{Network construction}\label{sec:appendix_network}

\subsubsection{Firm-technology bipartite network}

We begin the construction of the firm-technology bipartite network by calculating the relative number of patents that are filed by a firm $f$ and related to a technology $t$, which is normalized by the number of applicant firms and the number of related technologies of each patent.
Formally, we define the relative number of patents $X_{ft}$ as:
\begin{equation}
X_{ft}=\sum_{f \in m(p), t \in n(p)}\frac{1}{|m(p)||n(p)|}, \tag{B.1}
\end{equation}
where $m(p)$ is the set of applicant firms of patent $p$, and $n(p)$ is the set of technologies where patent $p$ relates.
To further decide whether technology $t$ is within the primary technologies of firm $f$, we utilize the Revealed Comparative Advantage (RCA) index~\citep{balassa1965trade}, which reads
\begin{equation}
RCA_{ft}=\frac{X_{ft}}{\sum_{ft'}X_{ft'}}/\frac{\sum_{f't}X_{f't}}{\sum_{f't'}X_{f't'}}. \tag{B.2}
\end{equation}
Based on this, we can build the binary firm-technology bipartite network, whose adjacency matrix $M$ is defined as:
\begin{equation}
M_{ft}=
\begin{cases}
1, & RCA_{ft} \geq 1\\
0, & RCA_{ft} < 1
\end{cases}. \tag{B.3}
\end{equation}
In the constructed firm-technology bipartite network, the link between firm $f$ and technology $t$ indicates that $t$ is one of the primary technologies of $f$.

\subsubsection{Industry-technology bipartite network}

Based on the constructed firm-technology bipartite network, we first utilize the Bipartite Configuration Model (BiCM)\footnote{\url{https://github.com/mat701/BiCM}}~\citep{saracco2015randomizing,vallarano2021fast} to generate $1,000$ randomized networks that preserve the degree distribution on average.
Then for both empirical and generated networks, we aggregate all firm-technology links into weighted industry-technology links according to firms' NAICS codes, leading to the empirical and generated industry-technology networks.
The weight of each industry-technology link represents the number of firms within that industry for which the technology is their primary technology.

To validate the statistical significance of each empirical industry-technology link~\citep{cimini2022meta}, we measure the p-value of each link as the frequency whose weight is smaller than the corresponding ones in $1,000$ generated networks.
To control the false discovery rate in multiple hypothesis testing, we follow the Benjamini-Hochberg procedure~\citep{benjamini1995controlling} to determine the adjusted p-value threshold.
Specifically, we choose the p-value significant level $p\leq0.05$ and the false discovery level $\alpha=0.05$, links that satisfy both criteria are regarded as significant.
These significant empirical industry-technology links reveal the related technologies of each industry, based on which we can build the validated industry-technology bipartite network and assess the degree of ``agreement'' of firms' patenting activities with respect to their industries.

\subsection{The Fitness-Complexity method}\label{sec:fc}

Originally introduced for the country-product network built from international trade data, the fitness score $F_f$ of firm $f$ is defined as the sum of the complexity scores of its primary technologies.
The complexity score $Q_t$ of technology $t$ is defined as a nonlinear function of the fitness scores of the firms that have $t$ as one of their primary technologies.
The iterative process reads~\citep{tacchella2012new}
\begin{equation}
\begin{cases}
\tilde{F}_f^{(N+1)}=\sum_{t}M_{ft}Q_{t}^{(N)}\\
\\
\tilde{Q}_t^{(N+1)}=\frac{1}{\sum_{t}{M_{ft}/F_{f}^{(N)}}}
\end{cases}, \tag{B.4}
\end{equation}
where $M_{ft}$ is the adjacency matrix of the firm-technology network.
After each iteration step, the fitness and complexity scores are normalized by their average scores:
\begin{equation}
\begin{cases}
F_f^{(N)}=\tilde{F}_f^{(N)}/\langle \tilde{F}_f^{(N)} \rangle_f\\
\\
Q_t^{(N)}=\tilde{Q}_t^{(N)}/\langle \tilde{Q}_t^{(N)}\rangle_t
\end{cases}. \tag{B.5}
\end{equation}
Based on the firm-technology network, we employ the Fitness-Complexity method to calculate the fitness scores of firms and the complexity scores of technologies.
Given the convergence issues of the original Fitness-Complexity method, we employ a non-homogeneous version~\citep{servedio2018new}.
The iterative process will be halted at step $n$ when both Spearman correlation coefficients between the rankings of firms and technologies at step $n$ and the rankings of firms and technologies at step $n-100$ exceed $1-10^{-6}$.

\subsection{Coherent technological diversification}\label{sec:ctd}

In this study, we introduce various measurements that utilize information from applicants, references, and classifications of patents to assess the proximity among technologies~\citep{yan2017measuring}.

The applicant-based measure defines the proximity between technologies $t$ and $t'$ as the minimum of the pairwise conditional probabilities that a firm has one technology as the primary technology, given that it has another technology as the primary technology~\citep{hidalgo2007product,zaccaria2014taxonomy,pugliese2019coherent,hidalgo2021economic}:
\begin{equation}
\varphi_{tt'} = \frac{1}{max(u_t, u_{t'})}\sum_{f}{M_{ft}M_{ft'}}\label{eq:B.6} \tag{B.6}
\end{equation}
or
\begin{equation}
\varphi_{tt'} = \frac{1}{max(u_t, u_{t'})}\sum_{f}\frac{{M_{ft}M_{ft'}}}{d_f},\label{eq:B.7} \tag{B.7}
\end{equation}
where $u_{t}$ is the number of firms having technology $t$ as the primary technology, $d_f$ is the number of primary technologies of firm $f$.
Based on equations~\ref{eq:B.6} and~\ref{eq:B.7}, we can build the technological space (TS) and technological relatedness (TN) network, respectively.

The reference-based measure defines the proximity between technologies $t$ and $t'$ as the Jaccard similarity of patents' references in two technologies~\citep{yan2017measuring}:
\begin{equation}
\varphi_{tt'} = \frac{|R_t \bigcap R_{t'}|}{|R_t \bigcup R_{t'}|}, \tag{B.8}
\end{equation}
where $R_t$ is the set of references cited by patents related to technology $t$.

The classification-based measure defines the proximity between technologies $t$ and $t'$ as the Jaccard similarity of patents related to two technologies~\citep{yan2017measuring}:
\begin{equation}
\varphi_{tt'} = \frac{|N_t \bigcap N_{t'}|}{|N_t \bigcup N_{t'}|}, \tag{B.9}
\end{equation}
where $N_t$ is the set of patents related to technology $t$.

According to the above definition of technological proximity, the coherence between technological $t$ and the technological portfolio of firm $f$ is measured as~\citep{pugliese2019coherent}:
\begin{equation}
\gamma_{ft} = \sum_{t'}\varphi_{tt'}M_{ft'},\label{eq:B.10} \tag{B.10}
\end{equation}
or
\begin{equation}
\gamma_{ft} = \sum_{t'}\frac{\varphi_{tt'}M_{ft'}}{d_f}.\label{eq:B.11} \tag{B.11}
\end{equation}
Equation~\ref{eq:B.10} measures the sum of proximity between a technology and a firm's technological portfolio, while equation~\ref{eq:B.11} calculates the mean proximity.

Furthermore, we measure the Coherent Technological Diversification (CTD) of a firm as the mean coherence of its primary technologies~\citep{pugliese2019coherent}:
\begin{equation}
\Gamma_f = \frac{\sum_{t}M_{ft}\gamma_{ft}}{d_f}. \tag{B.12}
\end{equation}
According to different combinations of technological proximity ($\varphi_{tt'}$) and coherence ($\gamma_{ft}$) computations, we can calculate $8$ kinds of CTD, as listed in Table~\ref{tab:4}.

\section{Details of the Gini importance} \label{sec:appendix_gini}

A random forest model consists of a large number of decision trees, each of which is built from a bootstrap sample of the training data.
The Gini impurity measures the probability of incorrectly classifying a randomly chosen sample if it were classified randomly according to the class distribution.
For a node $v$ with $N$ samples in a decision tree, the Gini impurity $G(v)$ is calculated as:
\begin{equation}
G(v) = 1 - \sum_{i=1}^C(\frac{N_{i,v}}{N})^2, \tag{C.1}
\end{equation}
where $N_{i,v}$ is the number of samples of class $i$ in node $v$, $C$ is the the number of classes.

When a feature is used to split a node in a decision tree, the goal is to reduce the Gini impurity in the resulting child nodes as much as possible.
For a node $v$ that splits based on a feature, we can calculate the Gini impurity of the left child node $v_L$ and the right child node $v_R$ as:
\begin{equation}
\begin{cases}
G(v_L) = 1 - \sum_{i=1}^C(\frac{N_{i,{v_L}}}{N_L})^2 \\ \\
G(v_R) = 1 - \sum_{i=1}^C(\frac{N_{i,{v_R}}}{N_R})^2 \\
\end{cases}. \tag{C.2}
\end{equation}
Then we can calculate the weighted average Gini impurity of the child nodes:
\begin{equation}
G_c(v) = \frac{N_L}{N}G(v_L) + \frac{N_R}{N}G(v_R). \tag{C.3}
\end{equation}
The reduction of the Gini impurity due to the split at node $v$ is:
\begin{equation}
RG(v) = G(v) - G_c(v). \tag{C.4}
\end{equation}
In a single decision tree $t$, the total reduction of the Gini impurity contributed by feature $f$ is the sum of the reductions by nodes that splits based on feature $f$, which can be normalized among all features:
\begin{equation}
NRG(t,f) = \frac{\sum_{v \in N(f)}RG(v)}{\sum_{f'}\sum_{v \in N(f')}RG(v)}, \tag{C.5}
\end{equation}
where $N(f)$ is the set of nodes that splits based on feature $f$.

Finally, the Gini importance of a feature can be quantified as its average effectiveness in reducing the Gini impurity across all trees within the forest.
The Gini importance of feature $f$ is calculated as:
\begin{equation}
GI(f) = \frac{1}{T}\sum_{t=1}^{T}NRG(t,f), \tag{C.6}
\end{equation}
where $T$ is the total number of decision trees.

\section{SHAP-based importance rankings of features} \label{sec:appendix_shap_importance}

\begin{figure*}[pos=hbt]
    \renewcommand{\thefigure}{D.1}
    \centering
    \caption{
    Rankings of the top $12$ features with the highest SHAP-based importance scores under different growth indicators.
    Each bar corresponds to a feature.
    The inset shows the importance rankings of all features.}
    \includegraphics[width=0.95\linewidth]{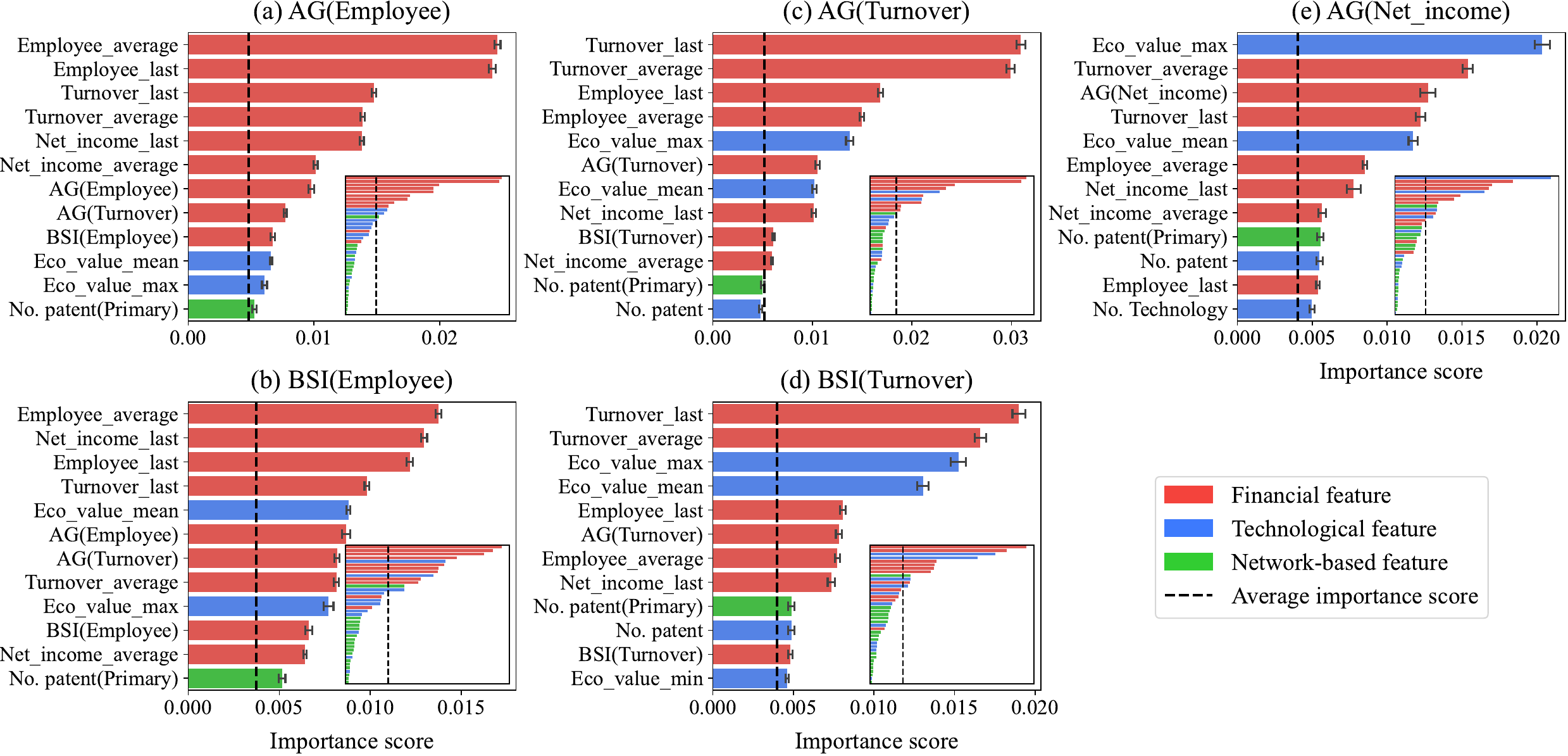}
    \label{fig:D.1}
\end{figure*}

Figure~\ref{fig:D.1} shows the rankings of the top $12$ features with the highest importance scores based on SHAP values, where the inset shows the rankings of all features.
The importance score of each feature is calculated as its average absolute SHAP value over all firms.
The rankings are similar with those presented in Fig.~\ref{fig:4} in the main text, which supports the robustness of our findings in Section~\ref{sec:feature_importance}.

\FloatBarrier
\section{Supplementary materials} \label{appendix:sm}
% The supplementary material related to this paper is available in the file \textit{Supplementary Material.pdf}.
The supplementary material related to this paper can be requested from Y.H. (hywei@std.uestc.edu.cn).
This document includes additional experiments conducted to verify the robustness and consistency of our findings.
It is structured as follows:
\begin{itemize}
    \item Section S1: Results of random forest models within different time windows
    \item Section S2: Results of XGBoost models within different time windows
    \item Section S3: Results under less restrictive data filtering criteria
\end{itemize}
These supplementary analyses support the primary findings discussed in the main text across different time windows, predictive models, and data filtering criteria.

\FloatBarrier
\bibliographystyle{apalike-doi}
\bibliography{reference}

\end{document}